\documentclass[onefignum,onetabnum]{siamonline250211}

\usepackage{lipsum}
\usepackage{amsfonts}
\usepackage{graphicx}
\usepackage{epstopdf}
\usepackage{algorithmic}
\usepackage{mathtools}
\usepackage{booktabs}
\usepackage{array}
\usepackage{needspace}
\usepackage{dsfont}
\usepackage{amsopn}
\usepackage{placeins}

\ifpdf
  \DeclareGraphicsExtensions{.eps,.pdf,.png,.jpg}
\else
  \DeclareGraphicsExtensions{.eps}
\fi

\usepackage{enumitem}
\setlist[enumerate]{leftmargin=.5in}
\setlist[itemize]{leftmargin=.5in}

\newsiamremark{remark}{Remark}
\crefname{remark}{Remark}{Remarks}
\newsiamremark{hypothesis}{Hypothesis}
\crefname{hypothesis}{Hypothesis}{Hypotheses}
\newsiamthm{claim}{Claim}
\newsiamremark{fact}{Fact}
\crefname{fact}{Fact}{Facts}

\newcommand{\bsx}{\boldsymbol{x}}
\newcommand{\bsz}{\boldsymbol{z}}
\newcommand{\bsy}{\boldsymbol{y}}
\newcommand{\bsV}{\boldsymbol{V}}
\newcommand{\bsD}{\boldsymbol{D}}
\newcommand{\bsX}{\boldsymbol{X}}
\newcommand{\bsY}{\boldsymbol{Y}}
\newcommand{\bsZ}{\boldsymbol{Z}}
\newcommand{\Lcal}{\mathcal{L}}
\newcommand{\bsXsf}{\boldsymbol{\mathsf{X}}}
\newcommand{\Xsf}{\mathsf{X}}
\newcommand{\EE}{\mathbb{E}}
\newcommand{\PP}{\mathbb{P}}
\newcommand{\VV}{\mathbb{V}}

\providecommand{\Lcal}{\mathcal{L}}

\DeclareMathOperator{\Cov}{Cov}
\DeclareMathOperator{\tr}{tr}

\mathtoolsset{showonlyrefs=false}
\allowdisplaybreaks

\headers{Coupled MCMC for Shapley effects}{M. Hirono and T. Goda}

\title{Coupled unbiased MCMC for estimating Shapley effects under dependent inputs}

\author{Masataka Hirono\thanks{Graduate School of Engineering, The University of Tokyo, 7-3-1 Hongo, Bunkyo-ku, Tokyo 113-8656, Japan  (\email{mskhirono@g.ecc.u-tokyo.ac.jp}).}
\and Takashi Goda\thanks{Graduate School of Information Science and Technology, The University of Tokyo, 7-3-1 Hongo, Bunkyo-ku, Tokyo 113-8656, Japan (\email{takashigoda@g.ecc.u-tokyo.ac.jp}).}
}

\ifpdf
\hypersetup{
  pdftitle={Coupled MCMC for Shapley effects},
  pdfauthor={M. Hirono and T. Goda}
}
\fi

\makeatletter
\let\arxivmaketitle\maketitle
\let\arxivinternalmaketitle\@maketitle
\let\arxivthanks\thanks
\let\bibcite\svbibcite
\makeatother

\begin{document}

\maketitle
\raggedbottom


\begin{abstract}
In global sensitivity analysis, Shapley effects provide a game-theoretic allocation of output variance among potentially dependent inputs. Their computation typically requires repeated conditional sampling, which can be challenging even when independent samples from the joint input distribution can be generated exactly. This difficulty arises naturally in hierarchical models, where conditioning on arbitrary subsets need not follow the generative order. To efficiently estimate Shapley effects under such dependence between input variables, we propose a permutation-based framework using coupled unbiased Markov chain Monte Carlo (MCMC). Our proposed method combines two complementary coupling mechanisms: (i) a joint coupling of conditional distributions along a random permutation, designed to reduce the variance of estimated marginal contributions, and (ii) a meeting-based coupling of Markov chains that removes finite-time MCMC bias through a telescoping correction. Under suitable conditions, we prove that the resulting estimators for the unnormalized Shapley effects are unbiased, and have finite variance and finite expected computational cost. Numerical experiments with various test cases illustrate the computational efficiency of the proposed method.
\end{abstract}

\begin{keywords}
variance-based sensitivity analysis, Shapley effects, dependent inputs,
conditional sampling, coupling, unbiased Markov chain Monte Carlo
\end{keywords}
\begin{MSCcodes}
65C05, 65C40, 62J10, 62H12
\end{MSCcodes}


\section{Introduction}\label{sec:introduction}

Global sensitivity analysis aims to quantify how uncertainty in the inputs of a mathematical or computational model contributes to uncertainty in its output. It provides a basis for identifying influential inputs, prioritizing uncertainty reduction, and interpreting the behavior of complex models. We refer to \cite{Primer08,DGIP21} for comprehensive accounts of global sensitivity analysis. Among the most widely used measures are the first-order and total Sobol' indices \cite{S93,S01}. For models with independent inputs, these indices have a natural interpretation through the functional analysis-of-variance decomposition. With dependent inputs, however, the effects of the model response and those of the input dependence are intertwined, which makes the interpretation of individual variance contributions more delicate \cite{IP19}.

Shapley effects provide an alternative by applying the Shapley value from cooperative game theory to conditional variance components \cite{O14}. The resulting unnormalized Shapley effects are nonnegative and sum exactly to the total output variance. Dividing by this variance thus gives normalized effects that sum to one. Importantly, these properties hold without assuming input independence \cite{SNS16,OP17}. Rather than separating response interactions from statistical dependence, Shapley effects allocate their combined contributions among the inputs \cite{IP19}. This makes them useful if input dependence is an essential feature of the stochastic model.

Computing Shapley effects involves two distinct difficulties. The first is combinatorial, as each Shapley effect is a weighted average of marginal increments over all subsets of the other inputs. Sampling uniformly random permutations avoids exhaustive enumeration and provides one increment for every input in each replication \cite{CGT09,SNS16}. The second difficulty is the evaluation of the conditional variance components entering those increments. Monte Carlo estimators typically require repeated sampling from conditional input distributions. Several approaches proposed in the literature \cite{BBD20,PRB21} reduce the variance or computational cost of these estimators, but efficient sampling from the joint input distribution does not by itself provide an efficient conditional sampler for every subset.

This difficulty arises naturally in hierarchical models. For example, uncertain distributional parameters may first be sampled and then used to generate a load or an environmental forcing. The joint input vector is readily generated in this order. In contrast, fixing a realized load and resampling its distributional parameters reverses the generative direction and may lead to a non-standard conditional distribution. Hierarchical models of this type can be found in reliability sensitivity analysis \cite{EPS24}. Since the estimation of Shapley effects involves arbitrary predecessor subsets, the required conditioning operations cannot be restricted to the natural generative order. Therefore, in this paper, we focus on settings where independent samples from a specified joint input distribution can be generated exactly, but a dedicated exact sampler for every required conditional distribution is not assumed available. These conditional distributions are instead explored using Markov chain Monte Carlo (MCMC) methods.

A related line of work constructs sensitivity index estimators or conditional samplers from available samples. Broto et al.\ \cite{BBD20} developed nearest-neighbor methods for unknown input distributions, while Wang \cite{W26} recently proposed to use generative models, including diffusion-based conditional sampling, for global sensitivity analysis of Bayesian-calibrated models. Our approach instead uses Markov kernels that leave the prescribed conditional distributions invariant. Its objective is to remove the finite-time bias of conditional MCMC without replacing the prescribed conditional laws with learned approximations.

Our starting point is the permutation-based pick-freeze estimator \cite{G21}. For models with independent inputs, it generates two independent input vectors and successively replaces coordinates of one vector by those of the other, following a random permutation. Every intermediate vector has the correct joint distribution, and adjacent vectors differ in only one coordinate. This permits both the reuse of model evaluations and direct estimation of adjacent differences using shared randomness. Under input dependence, however, coordinate exchange generally destroys the prescribed input distribution. The extension discussed in \cite[Section~5.2]{G21} replaces coordinate exchange by conditional resampling, but generating new conditional samples independently at successive steps can increase the variance of the estimators of the marginal contributions. We address this issue by \emph{coupling the conditional simulations across the whole permutation}.

The method we propose in this paper combines two complementary coupling mechanisms. \emph{Cross-target coupling} links conditional simulations along a random permutation as discussed above, and is designed to reduce the variance of adjacent differences. \emph{Within-target coupling} uses lagged replicas and telescoping corrections to remove finite-time MCMC bias \cite{JOA20}. Coupling both within and across targets is also used in multilevel unbiased estimation \cite{HJLT23}, which combines MCMC debiasing with randomization over levels to remove discretization error. Here, however, the levels correspond to successively released input coordinates, and every adjacent difference is assigned to a different input in the Shapley allocation. There is no discretization limit or infinite level expansion to debias.

By reusing corrected level estimates in neighboring increments, we obtain simultaneous estimators of all unnormalized Shapley effects. Under suitable conditions, we establish their unbiasedness, finite variance, and finite expected computational cost. We derive variance bounds that account for cross-level dependence, and show that the estimated contributions sum samplewise to an unbiased estimator of the total output variance. 

The remainder of this paper is organized as follows. \Cref{sec:preliminaries} introduces Shapley effects, their conditional pick-freeze representation, and unbiased MCMC. \Cref{sec:estimation} develops the coupled estimator, its variance bounds, and computational cost analysis. \Cref{sec:numerical} presents the numerical experiments for various test cases, and \cref{sec:conclusion} concludes this paper.

\section{Preliminaries}\label{sec:preliminaries}

Let $\bsX=(X_1,\dots,X_d)\sim \rho$, where each $X_j$ takes values in a standard Borel space $\Xsf_j$. We write $\bsXsf=\prod_{j=1}^d\Xsf_j$, equipped with the product $\sigma$-algebra. Regular conditional distributions then exist \cite{K21}. Let $f$ be a measurable real-valued response with $f\in L^2(\rho)$ and $\sigma^2 = \VV[f(\bsX)]>0$. We use the shorthand $[d]=\{1,\ldots,d\}$, and, for any subset $u\subseteq [d]$, let $\bsX_u=(X_j)_{j\in u}$ and $-u=[d]\setminus u$. For a probability measure $\mu$ and a $\mu$-integrable function $h$, we write $\mu(h)=\int h\, \mathrm{d} \mu.$

\subsection{Shapley effects}\label{subsec:shapley-effects}

For each $u\subseteq [d]$, we define the conditional variance component 
\[ C(u)=\EE\!\left[ \VV\!\left[ f(\bsX)\,|\, \bsX_{-u}\right] \right], \qquad u\subseteq [d]. \]
In particular, we have $C(\emptyset)=0$ and $C([d])=\sigma^2$. The unnormalized Shapley effect for the $j$-th input variable $X_j$ is given by
\begin{align}\label{eq:shapley_effect}
    \phi_j = \frac{1}{d}\sum_{u\subseteq -\{j\}}\binom{d-1}{|u|}^{-1}\left( C(u\cup\{j\})-C(u)\right).
\end{align}
These effects satisfy $0\le \phi_j\le \sigma^2$ and $\sum_{j=1}^{d}\phi_j=\sigma^2$. The normalized Shapley effects are denoted by $\tilde{\phi}_j = \phi_j/\sigma^2\in [0,1]$.  The Shapley allocation associated with $C$ coincides with that obtained from the explained-variance components $\VV\!\left[\EE\!\left[f(\bsX) \,|\, \bsX_u\right]\right]$; see \cite[Theorem~1]{SNS16}.

Let $\pi$ be a uniformly random permutation of $[d]$, independent of $\bsX$, and let $P_j(\pi)$ denote the set of indices preceding $j$ in $\pi$. An equivalent representation of
\eqref{eq:shapley_effect} is
\begin{align}\label{eq:shapley_effect_permutation}
    \phi_j = \EE_{\pi}\!\left[ C(P_j(\pi)\cup\{j\})-C(P_j(\pi))\right].
\end{align}
Set $U_0=\emptyset$ and $U_\ell=\{\pi(1),\ldots,\pi(\ell)\}$ for $\ell=1,\ldots,d$. Each permutation then supplies the increment $C(U_\ell)-C(U_{\ell-1})$ for the $\pi(\ell)$-th input variable $X_{\pi(\ell)}$. These increments satisfy the telescoping identity 
\[ \sum_{\ell=1}^d[C(U_\ell)-C(U_{\ell-1})]=C([d])-C(\emptyset)=\sigma^2.\]

\subsection{Conditional pick-freeze representation}
\label{sec:conditional-identity}

Fix $u\subseteq[d]$. Given an outer draw $\bsX\sim\rho$, generate a conditional copy $\bsZ$ with law
\[ \Lcal(\bsZ\,|\,\bsX=\bsx) = \Lcal_\rho(\bsX\,|\,\bsX_{-u}=\bsx_{-u}). \]
The conditional law of $\bsZ$ depends on $\bsX$ only through $\bsX_{-u}$. Therefore, $f(\bsX)$ and $f(\bsZ)$ are conditionally independent and identically distributed given $\bsX_{-u}$. Expanding the square gives the conditional pick-freeze identity (see, e.g., \cite[Section~5.2]{G21})
\begin{equation}\label{eq:pickfreeze}
 \EE\!\left[
 \frac{(f(\bsX)-f(\bsZ))^2}{2}
 \,\middle|\,\bsX_{-u}\right]
 =
 \EE\!\left[ (f(\bsX))^2 \,|\, \bsX_{-u}\right]-\left(\EE\!\left[ f(\bsX) \,|\, \bsX_{-u}\right]\right)^2
 =
 \VV[f(\bsX) \,|\,\bsX_{-u}].
\end{equation}
Both sides are integrable because $f\in L^2(\rho)$, and their expectation is $C(u)$. 

For the permutation prefixes $U_{\ell}$, let us define the full-vector conditional distributions 
\[  \mu_\ell^{\bsx} = \mathcal{L}_{\rho}\!\left(\bsX\,|\, \bsX_{-U_\ell}=\bsx_{-U_\ell} \right), \qquad \ell=0,\ldots,d,
\]
for $\rho$-almost every $\bsx$. The coordinates in $-U_{\ell}$ are fixed at $\bsx_{-U_{\ell}}$, and $\mu_0^{\bsx}=\delta_{\bsx}$ and $\mu_d^{\bsx}=\rho$. Notice that the dependence on $\pi$ is suppressed in the notation here. Writing $\psi_{\bsx}(\bsz)=(f(\bsx)-f(\bsz))^2/2$, the above identity \eqref{eq:pickfreeze} implies
\[
 \EE\!\left[
 \mu_\ell^{\bsX}(\psi_{\bsX})\,|\, \pi
 \right]=C(U_\ell),
 \qquad \ell=0,\ldots,d.
\]
Here the outer expectation is essential:
the inner quantity $\mu_\ell^{\bsx}(\psi_{\bsx})$ averages only over the conditional
copy at a fixed $\bsx$.

Suppose that, conditional on $(\bsX,\pi)$, the joint distribution of $(\bsZ_0,\ldots,\bsZ_d)$ has marginals $\mu_0^{\bsX},\ldots,\mu_d^{\bsX}$. Then
\[ \EE\!\left[ \psi_{\bsX}(\bsZ_{\ell})-\psi_{\bsX}(\bsZ_{\ell-1})\,|\, \pi\right] = C(U_\ell)-C(U_{\ell-1}). \]
Thus, the conditional samples at different levels need not be independent: their marginal laws determine the expectations, whereas their coupling affects the variance of adjacent differences. For independent inputs, a joint construction can be simply given by $\bsZ_{\ell}=(\bsY_{U_{\ell}}, \bsX_{-U_{\ell}})$, where a single $\bsY\sim \rho$ is independent of $(\bsX,\pi)$. This gives the original coordinate-swapping construction of \cite{G21}.

\subsection{Unbiased Markov chain Monte Carlo}\label{sec:unbiased-mcmc}

We recall the meeting-based unbiased MCMC correction \cite{GR14,JOA20}, which will be applied to the conditional expectations above. We refer to \cite{AJ26} for an introductory account and \cite{MDDJ20} for some extensions.

Let $K$ be a Markov kernel with invariant probability law $\mu$, and suppose that $\mu(|h|)<\infty$. Consider two chains $(\bsZ_t)_{t\ge0}$ and $(\widetilde{\bsZ}_t)_{t\ge0}$ with the same initial marginal law and transition kernel $K$. Advance the first chain once, then couple the updates of $(\bsZ_t,\widetilde{\bsZ}_{t-1})$ while preserving both marginal transitions. Define the meeting time
\[
 \tau=\inf\left\{t\ge1 \,|\, \bsZ_t=\widetilde{\bsZ}_{t-1}\right\}.
\]
Assume that $\tau<\infty$ almost surely and that the coupling is \emph{faithful}, meaning that $\bsZ_t=\widetilde{\bsZ}_{t-1}$ for every $t\ge\tau$.

Fix an integer $k\ge1$ and assume that $h(\bsZ_t)$ is integrable for every $t\ge k$. Since the two chains have the same marginal law at each time, we have
\begin{equation}\label{eq:telescoping-identity}
 \EE\!\left[h(\bsZ_t)-h(\widetilde{\bsZ}_{t-1})\right]
 =
 \EE\!\left[ h(\bsZ_t)\right]-\EE\!\left[ h(\bsZ_{t-1})\right],
 \qquad t\ge k+1.
\end{equation}
The corrected estimator of $\mu(h)$, starting at time $k$, is
\begin{equation}\label{eq:single-target-correction}
 H_k = h(\bsZ_k)+\sum_{t=k+1}^{\tau-1}
 \left[h(\bsZ_t)-h(\widetilde{\bsZ}_{t-1})\right],
\end{equation}
where an empty sum is understood to be zero.

\begin{lemma}[Unbiased MCMC correction]\label{lem:single-target}
Under the preceding assumptions, suppose that $\EE[h(\bsZ_t)] \to\mu(h)$ as $t\to \infty$ and
\begin{equation}\label{eq:single-target-integrability}
 \EE\!\left[|h(\bsZ_k)|\right]
 +\sum_{t=k+1}^{\infty}
 \EE\!\left[|h(\bsZ_t)-h(\widetilde{\bsZ}_{t-1})|\right]<\infty.
\end{equation}
Then $H_k$ given in \eqref{eq:single-target-correction} is integrable and has expectation $\mu(h)$.
\end{lemma}
\begin{proof}
By \eqref{eq:telescoping-identity}, the estimator with its correction sum truncated at a deterministic time $T>k$ has expectation $\EE[h(\bsZ_T)]$. Condition \eqref{eq:single-target-integrability} implies the convergence of these truncated estimators in $L^1$ as $T\to \infty$. Faithfulness makes every summand at times $t\ge\tau$ zero, so the limit is $H_k$. Thus, the assumed convergence of expectations gives $\EE[H_k]=\mu(h)$.
\end{proof}

\section{Coupled estimation of Shapley effects}\label{sec:estimation}

Following the discussion in \cref{sec:conditional-identity}, we aim to construct jointly unbiased estimates of the conditional expectations $\mu_1^{\bsx}(\psi_{\bsx}),\ldots,\mu_d^{\bsx}(\psi_{\bsx})$ for each outer pair $(\bsx,\pi)$. These estimates are reused in adjacent differences to obtain all input contributions. We first describe the coupled simulation and the resulting estimator, and then establish unbiasedness and analyze variance and computational cost.

Throughout the construction below, $\ell$ indexes the permutation prefixes and $t$ denotes MCMC time, while $(\bsx,\pi)$ remains fixed.

\subsection{Coupling across conditional targets}
\label{sec:coupled-estimator}\label{sec:transitions}

At level $\ell\in [d]$, the coordinates in $-U_\ell$ remain fixed at $\bsx_{-U_\ell}$. Define the conditional state space
\[ \bsXsf_{\ell}^{\bsx} = \left\{ \bsz\in \bsXsf \,|\, \bsz_{-U_{\ell}} = \bsx_{-U_{\ell}} \right\}, \]
equipped with the induced $\sigma$-algebra. Let $K_\ell$ be a Markov kernel on $\bsXsf_{\ell}^{\bsx}$ with invariant probability law $\mu_\ell^{\bsx}$. Let $\bsV$ be a random block with a fixed distribution independent of the current states. Suppose that a measurable update map $\Phi_\ell$ satisfies
\begin{equation*}
 \Lcal\{\Phi_\ell(\bsz,\bsV)\}=K_\ell(\bsz,\cdot),
 \qquad \bsz_{-U_\ell}=\bsx_{-U_\ell},
\end{equation*}
where the law on the left-hand side is taken over the randomness in $\bsV$, with $\bsz$ and $(\bsx,\pi)$ fixed. Equivalently, for every $\bsz\in \bsXsf_{\ell}^{\bsx}$ and every measurable set $A\subseteq \bsXsf_{\ell}^{\bsx}$, 
\[ \PP\!\left[ \Phi_{\ell}(\bsz,\bsV)\in A\right] = K_{\ell}(\bsz,A). \]
Each update depends only on its own state, the random block, the outer input, and the set $U_{\ell}$. We assume joint measurability in these arguments.

Start every level at $\bsZ_{\ell,0}=\bsx$ and generate independent copies $\bsV_0,\bsV_1,\ldots$ of $\bsV$, independently of the outer pair $(\bsx,\pi)$. Updating all levels by
\[
 \bsZ_{\ell,t+1}=\Phi_\ell(\bsZ_{\ell,t},\bsV_t),
 \qquad \ell=1,\ldots,d,
\]
preserves the marginal transition kernel $K_\ell$ at each level. The key ingredient is that sharing $\bsV_t$ couples the simulations across the different conditional targets.

\subsection{Coupling with lagged replicas}\label{sec:lagged-coupling}
For each level $\ell\in [d]$, we introduce an auxiliary chain with $\widetilde{\bsZ}_{\ell,0}=\bsx$. Advance the main chains once, $\bsZ_{\ell,1}=\Phi_\ell(\bsx,\bsV_0)$, and then update 
\begin{equation}\label{eq:lagged-updates}
 \bsZ_{\ell,t+1}=\Phi_\ell(\bsZ_{\ell,t},\bsV_t),
 \qquad
 \widetilde{\bsZ}_{\ell,t}
 =\Phi_\ell(\widetilde{\bsZ}_{\ell,t-1},\bsV_t),
 \qquad t\ge1.
\end{equation}
Note that, at each update, the same block is used across levels and within each main--auxiliary pair. Sharing the same block specifies the dependence between conditional simulations across levels, and also couples two chains with the same target at a one-step lag within each level.

Conditional on $(\bsx,\pi)$, the vectors $(\bsZ_{\ell,t})_{\ell=1}^d$ and $(\widetilde{\bsZ}_{\ell,t})_{\ell=1}^d$ have the same law for each $t$. The main vector uses $(\bsV_0,\ldots,\bsV_{t-1})$, while the auxiliary vector uses $(\bsV_1,\ldots,\bsV_t)$. These block sequences have the same probability law. This equal-time identity gives the
telescoping expectations in \eqref{eq:telescoping-identity}. Define the levelwise meeting times by
\begin{equation*}
 \tau_\ell=\inf\left\{t\ge1 \,|\,
       \bsZ_{\ell,t}=\widetilde{\bsZ}_{\ell,t-1}\right\},
\end{equation*}
with the convention $\inf \emptyset = \infty$. After a pair meets, applying the same map to equal states with the
same random block preserves equality:
\begin{equation*}
 \bsZ_{\ell,t}=\widetilde{\bsZ}_{\ell,t-1},
 \qquad t\ge\tau_\ell.
\end{equation*}
Thus the coupling is faithful at each level, and the corresponding correction can stop at its own meeting time.  However, almost sure finiteness of the meeting times and convergence to the target expectations do not follow from shared randomness alone; they require additional conditions on the kernels and update maps.

\begin{remark}[Construction of meeting couplings]
    Strategies for constructing meeting couplings include maximal couplings of proposals or conditional updates, reflection-based constructions, and combinations of contractive and meeting-producing updates; see \cite[Section~17.4]{AJ26} and \cite[Section~4]{JOA20}. General two-chain couplings can be represented by a joint update $(\bsz',\widetilde \bsz')=\Psi_\ell(\bsz,\widetilde \bsz,V)$, whose output marginals are $K_\ell(\bsz,\cdot)$ and $K_\ell(\widetilde \bsz,\cdot)$, respectively, for each fixed pair $(\bsz,\widetilde \bsz)$. Here we focus on couplings admitting the shared-map representation \eqref{eq:lagged-updates}. Sufficient conditions for geometric meeting-time bounds, based on drift conditions and a uniformly positive meeting probability on a suitable small set, are given in \cite[Section~3.2]{JOA20}; extensions under polynomial drift conditions are developed in \cite{MDDJ20}. For the examples in \cref{sec:numerical}, we implement Gibbs and independence Metropolis updates in the shared-map form \eqref{eq:lagged-updates} and verify the required convergence and meeting-time properties.
\end{remark}

\subsection{Contribution estimates}\label{sec:corrected}
Fix integers $1\le k\le m$ and put $n=m-k+1$. At each level $\ell\in [d]$, we average the single-start estimator in \eqref{eq:single-target-correction}, with $h=\psi_{\bsx}$, over the starting indices $k,\ldots,m$, using the same pair
of coupled trajectories. Collecting the base terms and correction terms gives
\begin{align}\label{eq:A}
 A_\ell = \frac1n\sum_{t=k}^m\psi_{\bsx}(\bsZ_{\ell,t})+\sum_{t=k+1}^{\tau_\ell-1}
 \min\left\{1,\frac{t-k}{n}\right\}
 \left[\psi_{\bsx}(\bsZ_{\ell,t})
       -\psi_{\bsx}(\widetilde{\bsZ}_{\ell,t-1})\right].
\end{align}
The weight of the correction at time $t$ is the fraction of these indices that are smaller than $t$. Set $A_0=0$ and reuse each interior $A_\ell$ in its two neighboring contributions:
\begin{equation}\label{eq:coupled-contributions}
 D_{\pi(\ell)}=A_\ell-A_{\ell-1},
 \qquad \ell=1,\ldots,d,
 \qquad
 \sum_{j=1}^dD_j=A_d.
\end{equation}

\Cref{alg:estimator} evaluates \eqref{eq:A} one update at a time. The variables $\bsz_\ell$ and $\bsy_\ell$ store the current main and auxiliary states, respectively, while $A_\ell$ is accumulated online. The auxiliary chain stops at meeting, while the main chain continues to $m$ if needed, so level $\ell$ is simulated through $\max(m,\tau_\ell)$.

\begin{algorithm}[h]
\caption{One vector of unbiased Shapley contributions}
\label{alg:estimator}
\begin{algorithmic}[1]
\REQUIRE A sampler for $\rho$, conditional maps $\Phi_\ell$,
         and integers $1\le k\le m$.
\ENSURE Contribution vector $(D_1,\ldots,D_d)$.
\STATE Draw $\bsx\sim\rho$ and an independent uniform permutation
       $\pi$; form $U_\ell$ and $\Phi_\ell$.
\STATE Evaluate $f(\bsx)$; set $n\gets m-k+1$, $t\gets0$,
       and $A_0\gets0$.
\STATE Set $(\bsz_\ell,\bsy_\ell,A_\ell,\tau_\ell)
       \gets(\bsx,\bsx,0,\infty)$ for all $\ell\in[d]$.
\WHILE{$t<m$ or some $\tau_\ell=\infty$}
 \STATE $t\gets t+1$; generate one fresh block $\bsV_{t-1}$.
 \FOR{$\ell\in[d]$ with $t\le m$ or $\tau_\ell=\infty$}
  \STATE $\bsz_\ell\gets\Phi_\ell(\bsz_\ell,\bsV_{t-1})$.
  \STATE \textbf{if} $t\ge2$ and $\tau_\ell=\infty$
         \textbf{then}
         $\bsy_\ell\gets\Phi_\ell(\bsy_\ell,\bsV_{t-1})$.
  \STATE \textbf{if} $\tau_\ell=\infty$ and $\bsz_\ell=\bsy_\ell$
         \textbf{then} $\tau_\ell\gets t$.
  \STATE \textbf{if} $k\le t\le m$
         \textbf{then} $A_\ell\gets A_\ell+\psi_{\bsx}(\bsz_\ell)/n$.
  \STATE \textbf{if} $t>k$ and $\tau_\ell=\infty$
         \textbf{then} $A_\ell\gets A_\ell+
          \min\{1,(t-k)/n\}
          [\psi_{\bsx}(\bsz_\ell)-\psi_{\bsx}(\bsy_\ell)]$
 \ENDFOR
\ENDWHILE
\STATE Set $D_{\pi(\ell)}\gets A_\ell-A_{\ell-1}$
       for all $\ell\in[d]$.
\RETURN $(D_1,\ldots,D_d)$.
\end{algorithmic}
\end{algorithm}

For a fixed integer $N\ge1$, run \cref{alg:estimator} independently $N$ times, using a fresh outer input, permutation, and sequence of random blocks in each run. Let $D_j^{(r)}$ and $A_d^{(r)}$ denote the corresponding quantities from the $r$-th run, and define our estimators
\begin{equation}\label{eq:outer-estimator}
 \widehat\phi_{j,N}=\frac1N\sum_{r=1}^ND_j^{(r)},
 \qquad
 \widehat\sigma_N^2=\sum_{j=1}^d\widehat\phi_{j,N}
                 =\frac1N\sum_{r=1}^NA_d^{(r)},
\end{equation}
for the unnormalized Shapley effects $\phi_j$ and the total output variance $\sigma^2$. Individual estimates $D_j$ may be negative, even though the true effects $\phi_j$ are nonnegative.

\subsection{Unbiasedness}\label{sec:unbiasedness}
For the proofs, all chains in \eqref{eq:lagged-updates} are continued indefinitely with the same sequence of random blocks, even after \cref{alg:estimator} stops storing a chain. Faithfulness ensures that this continuation does not change the estimator. We first establish unbiasedness of our estimators under convergence and integrability assumptions. Sufficient moment and meeting-time conditions for the integrability requirement are given later in \cref{sec:variance}.

\begin{theorem}[Unbiased contribution vector]\label{thm:unbiased}
Suppose that the coupling in \eqref{eq:lagged-updates} is faithful and every $\tau_\ell$ is almost surely finite. For almost every $(\bsx,\pi)$, assume at every level that
\begin{equation}\label{eq:mean-convergence}
 \EE[\psi_{\bsx}(\bsZ_{\ell,t})\,|\,\bsx,\pi]
 \longrightarrow\mu_\ell^{\bsx}(\psi_{\bsx}),
 \qquad t\to\infty.
\end{equation}
In addition, assume for every level that
\begin{equation}\label{eq:L1-assumption}
 \EE\!\left[
 \frac1n\sum_{t=k}^m\left|\psi_{\bsX}(\bsZ_{\ell,t})\right|
 +\sum_{t=k+1}^{\infty}\min\left\{1,\frac{t-k}{n}\right\}
 \left|\psi_{\bsX}(\bsZ_{\ell,t})-
          \psi_{\bsX}(\widetilde{\bsZ}_{\ell,t-1})\right|
 \right] < \infty.
\end{equation}
Then \cref{alg:estimator} terminates almost surely and returns integrable variables satisfying
\begin{equation*}
 \EE\! \left[A_\ell\,|\, \bsX,\pi\right]=\mu_\ell^{\bsX}(\psi_{\bsX}),
 \qquad \EE[D_j]=\phi_j,
 \qquad \sum_{j=1}^dD_j=A_d,
 \qquad \EE[A_d]=\sigma^2.
\end{equation*}
\end{theorem}
\begin{proof}
The unconditional integrability assumption implies its conditional counterpart for almost every outer pair $(\bsx,\pi)$. Fix such a pair. For every $t>k$, the weight
$\min\{1,(t-k)/n\}$ is at least $1/n$. Thus, for each starting index $s\in\{k,\ldots,m\}$, the expected sum of the absolute unweighted corrections after $s$ is at most $n$ times the corresponding weighted sum in \eqref{eq:L1-assumption}. Every base value in this window is integrable. Since $\widetilde{\bsZ}_{\ell,t-1}$ and $\bsZ_{\ell,t-1}$ have the same conditional law, the triangle inequality and the integrable correction differences imply, by induction, that $\psi_{\bsx}(\bsZ_{\ell,t})$ is integrable for every $t\ge k$. Applying \cref{lem:single-target} with $h=\psi_{\bsx}$ therefore gives expectation $\mu_\ell^{\bsx}(\psi_{\bsx})$ for every single-start estimator. Their average is \eqref{eq:A}.

Integration over $\bsX$ gives $C(U_\ell)$ conditional on $\pi$, by \eqref{eq:pickfreeze}. Averaging the adjacent increment assigned to input $j$ over $\pi$ then gives \eqref{eq:shapley_effect_permutation}. Reusing each level value yields $\sum_jD_j=A_d$. The assumed absolute integrability justifies these integrations. Finally, the simulation terminates by $\max(m,\tau_1,\ldots,\tau_d)$, which is finite almost surely.
\end{proof}

\begin{corollary}[Recovery of coordinate swapping]\label{cor:independent}
Suppose that the input variables are mutually independent. At each update, draw a fresh $\bsY\sim\rho$, independently of the past and the outer input, and use $\Phi_\ell(\bsz,\bsY)=(\bsY_{U_\ell},\bsx_{-U_\ell})$ at every level and for both chains. If $k=m=1$, \cref{alg:estimator} returns the coordinate-swapping estimator of \cite{G21} sample by sample.
\end{corollary}

\begin{proof}
The first common draw gives $\bsZ_{\ell,1}=(\bsY_{U_\ell},\bsx_{-U_\ell})$ at every level. The next update maps both members of each lagged pair to the same state, so $\tau_\ell\le2$. For $k=m=1$, the correction in \eqref{eq:A} is empty and $A_\ell=\psi_{\bsx}((\bsY_{U_\ell},\bsx_{-U_\ell}))$. The adjacent differences in \eqref{eq:coupled-contributions} therefore coincide with the coordinate-swapping estimator.
\end{proof}

\begin{remark}[Augmented-state implementations]\label[remark]{rem:augmented-state}
Suppose that a level is simulated on an augmented space with invariant law $\nu_\ell^{\bsx}$ and a measurable map $\mathcal B$ satisfying $\nu_\ell^{\bsx}\circ\mathcal B^{-1}=\mu_\ell^{\bsx}$. Apply \cref{lem:single-target} on that space with test function $\psi_{\bsx}\circ\mathcal B$, and set $\bsZ_{\ell,t}=\mathcal B(\boldsymbol W_{\ell,t})$ for response evaluation. Start the main and auxiliary augmented chains at the same state at each level, with image $\bsx$, and use the same joint update construction. Meeting is checked on the augmented states. The equal-time joint laws and faithfulness then hold for the projected response evaluations, and
\[
 \nu_\ell^{\bsx}(\psi_{\bsx}\circ\mathcal B)
 =\mu_\ell^{\bsx}(\psi_{\bsx}).
\]
Under the corresponding convergence and integrability conditions, the unbiasedness proof above applies to the augmented chains. The moment bounds below also apply to the projected response evaluations under their respective moment and meeting-time assumptions. The projected process need not itself be Markov.
\end{remark}

\subsection{Variance bounds}\label{sec:variance}
We study how coupling across levels affects the variance and give sufficient conditions for the corrected estimators to have finite second moments. We first compare shared and conditionally independent level estimates, then bound the adjacent differences and their meeting corrections. We then obtain unconditional finite-variance guarantees, examine variation in an independent input, and describe standard error estimates.

\begin{proposition}[Comparison with independent levels]
\label{prop:independent-level-comparison}
Suppose that $A_1,\ldots,A_d$ have finite second moments. Conditional on $(\bsX,\pi)$, let $A_1^{\mathrm{ind}},\ldots,A_d^{\mathrm{ind}}$ be independent with
\[
 \Lcal(A_\ell^{\mathrm{ind}} \,|\, \bsX,\pi)
 = \Lcal(A_\ell \,|\, \bsX,\pi),\qquad \ell=1,\ldots,d.
\]
Set $A_0^{\mathrm{ind}}=0$ and $D_{\pi(\ell)}^{\mathrm{ind}} =A_\ell^{\mathrm{ind}}-A_{\ell-1}^{\mathrm{ind}}$. Then
\begin{equation}\label{eq:trace-gain}
 \sum_{j=1}^d\VV[D_j^{\mathrm{ind}}]
 -\sum_{j=1}^d\VV[D_j]
 =2\EE\!\left[
   \sum_{\ell=2}^d
   \Cov(A_\ell,A_{\ell-1}\,|\,\bsX,\pi)\right].
\end{equation}
Moreover, $\sum_{j=1}^dD_j^{\mathrm{ind}}$ and $\sum_{j=1}^dD_j$ have the same conditional distribution given $(\bsX,\pi)$.
\end{proposition}

\begin{proof}
The conditional means and variances of the individual level estimates are unchanged. For each fixed outer pair, expanding the variance of an adjacent difference gives
\[ \VV[D_{\pi(\ell)}^{\mathrm{ind}}\,|\, \bsX,\pi] -\VV[D_{\pi(\ell)}\,|\, \bsX,\pi] = 2\Cov(A_\ell,A_{\ell-1}\,|\, \bsX,\pi).\]
The conditional means of the contributions $D_j$ and $D_j^{\mathrm{ind}}$ also agree. The law of
total variance and summation over the inputs give \eqref{eq:trace-gain}. Finally, the two contribution sums telescope to $A_d^{\mathrm{ind}}$ and $A_d$, whose conditional laws agree.
\end{proof}

The two sums of variances in \eqref{eq:trace-gain} are the traces of the contribution-vector covariance matrices. Thus, positive average covariance between adjacent corrected level estimates reduces the trace relative to conditionally independent levels. Sharing randomness alone does not ensure the sign of this covariance. Since the law of the total estimate is unchanged, the improvement concerns the allocation among inputs rather than estimation of the total variance. We refer to \cref{app:correction-comparison} in the supplementary material for a sufficient condition for variance reduction of the uncorrected averages to persist after correction.

\paragraph{Adjacent differences}\label{sec:correction-moments}
Fix an outer pair $(\bsx,\pi)$. In the following conditional bounds, all probabilities, expectations, covariances, and norms are conditional on this pair. We use the indefinitely continued chains described in \cref{sec:unbiasedness} and define
\[
 G_{\ell,t}=\psi_{\bsx}(\bsZ_{\ell,t})-
                  \psi_{\bsx}(\bsZ_{\ell-1,t}),
 \qquad \bsZ_{0,t}=\bsx.
\]
Define $\widetilde G_{\ell,t}$ in the same way from the auxiliary chains, with
$\widetilde{\bsZ}_{0,t}=\bsx$. Set $\tau_0=1$ and
$T_\ell=\max(\tau_{\ell-1},\tau_\ell)$. Subtracting the two level estimates in
\eqref{eq:A} gives
\begin{equation}\label{eq:D-decomposition}
 D_{\pi(\ell)}=\frac1n\sum_{t=k}^m G_{\ell,t}+R_\ell,
 \qquad
 R_\ell=\sum_{s=1}^{\infty}\min\left(1,\frac{s}{n}\right)
       \left[G_{\ell,k+s}-\widetilde G_{\ell,k+s-1}\right].
\end{equation}
Faithfulness makes each summand in $R_\ell$ zero whenever $k+s\ge T_\ell$. The indefinite continuation makes this identity valid even when the algorithm has stopped storing one of the chains.

For $r\ge1$, write $\|Y\|_r=(\EE|Y|^r)^{1/r}$ for the conditional $L^r$ norm, with $\|Y\|_\infty$ denoting the conditional essential supremum. Assume that there exists  $p\in(2,\infty]$ such that
\begin{equation}\label{eq:contrast-moment}
 L_\ell=\sup_{t\ge0}\|G_{\ell,t}\|_p<\infty,
\end{equation}
for all $\ell\in [d]$, where $1/p=0$ for $p=\infty$. The equal-time joint laws established in \cref{sec:lagged-coupling} give the same moment bound for $\widetilde G_{\ell,t}$. Define the weighted meeting-tail sum
\begin{equation}\label{eq:tail-sum}
 b_\ell=\sum_{s=1}^{\infty}\min\left(1,\frac{s}{n}\right)
                  \PP(T_\ell>k+s)^\alpha, 
        \qquad \alpha=\frac12-\frac1p,
\end{equation}
The factor $L_\ell$ measures the size of adjacent response differences, while $b_\ell$ accounts for the meeting times and correction weights. A short meeting time alone need not produce a low-variance contribution when the adjacent response differences are large.

\begin{proposition}[Moments of an adjacent contribution]\label{prop:adjacent-moments}
Suppose that \eqref{eq:contrast-moment} holds and $b_\ell<\infty$. Then, the sum of the
absolute correction terms in \eqref{eq:D-decomposition} has conditional $L^2$ norm at most $2L_\ell b_\ell$. In particular, we have
\begin{equation}\label{eq:correction-bound}
 \|R_\ell\|_2\le 2L_\ell b_\ell,
 \qquad
 \|D_{\pi(\ell)}\|_2\le L_\ell(1+2b_\ell).
\end{equation}
The conditional variance also satisfies
\begin{equation}\label{eq:variance-bound}
 \sqrt{\VV[D_{\pi(\ell)}\,|\,\bsx,\pi]}
 \le
 \left\{\frac1{n^2}\sum_{s=k}^m\sum_{t=k}^m
             \Cov(G_{\ell,s},G_{\ell,t})\right\}^{1/2}
       +2L_\ell b_\ell.
\end{equation}
\end{proposition}

\begin{proof}
The difference $G_{\ell,k+s}-\widetilde G_{\ell,k+s-1}$ has $L^p$ norm at most $2L_\ell$ and vanishes whenever $k+s\ge T_\ell$. For $p<\infty$, H\"older's inequality therefore gives
\begin{align*}
 \|G_{\ell,k+s}-\widetilde G_{\ell,k+s-1}\|_2
 & = \left(\EE\!\left[ \left|G_{\ell,k+s}-\widetilde G_{\ell,k+s-1}\right|^2\mathds{1}_{T_{\ell}>k+s} \right]\right)^{1/2} \\
 & \le \left(\EE\!\left[ \left|G_{\ell,k+s}-\widetilde G_{\ell,k+s-1}\right|^p\right]\right)^{1/p} \PP\left(T_{\ell}>k+s\right)^{1/2-1/p} \\
 & \le 2L_\ell\PP(T_\ell>k+s)^\alpha.
\end{align*}
The same estimate follows from the essential supremum bound when $p=\infty$. Applying Minkowski's inequality to finite weighted sums of the absolute differences, then using monotone convergence, gives the stated bound on their absolute sum. It also bounds $\|R_\ell\|_2$. The time average in \eqref{eq:D-decomposition} has $L^2$ norm at most $L_\ell$, proving the second bound in \eqref{eq:correction-bound}. Finally, apply the triangle inequality to the centered terms in \eqref{eq:D-decomposition}. The variance of the time average is the double covariance sum in \eqref{eq:variance-bound}, while $\VV[R_\ell]\le\EE[R_\ell^2]$.
\end{proof}

The proof applies the moment and meeting-time argument of \cite{JOA20,MDDJ20} to adjacent differences rather than individual level estimates. The variance of the time average in \eqref{eq:variance-bound} is bounded by
\[
 \frac1n\sup_{s\ge k}\left\{
 \VV[G_{\ell,s}]+2\sum_{t>s}|\Cov(G_{\ell,s},G_{\ell,t})|\right\}.
\]
For fixed $k$, this gives an $O(n^{-1})$ bound for the variance of the time average when the supremum is finite and independent of $m$. The covariance condition controls the first term in \eqref{eq:variance-bound}, separately from the meeting correction.

If a conditional geometric bound
\[
 \PP(T_\ell>t)\le a_\ell q_\ell^{t-1},
 \qquad a_\ell\ge0,\quad 0<q_\ell<1,
\]
is available, then summing the weighted geometric series yields
\begin{equation}\label{eq:geometric-tail-sum}
 b_\ell\le
 a_\ell^\alpha q_\ell^{\alpha k}
 \frac{1-q_\ell^{\alpha n}}{n(1-q_\ell^\alpha)^2}.
\end{equation}
Indeed, $\sum_{s\ge1}\min\{1,s/n\}r^{s-1}=(1-r^n)/\{n(1-r)^2\}$ for $0<r<1$. Individual level-tail estimates give an adjacent bound through
\[
 \PP(T_\ell>t)\le
 \sum_{i\in\{\ell-1,\ell\}\cap[d]}\PP(\tau_i>t), \qquad t\ge1,
\]
without requiring independence of the meeting times.
For fixed $a_\ell,q_\ell,p$, and $k$, \eqref{eq:geometric-tail-sum} gives $b_\ell=O(n^{-1})$.

The moment factor $L_\ell$ can also be related directly to differences of model responses. The factorization
\begin{equation}\label{eq:factorization}
 \psi_{\bsx}(\bsz)-\psi_{\bsx}(\bsy)
 =\left\{f(\bsx)-\frac{f(\bsz)+f(\bsy)}2\right\}
       \{f(\bsy)-f(\bsz)\}
\end{equation}
implies that, if $f$ has finite range width
$\omega_f=\sup f-\inf f$, then 
\[
 L_\ell\le \min\left\{\frac{\omega_f^2}{2},\,
       \omega_f\sup_t\|f(\bsZ_{\ell,t})-f(\bsZ_{\ell-1,t})\|_p\right\}.
\]
Thus, a coupling that keeps adjacent responses close can reduce the moment factor bound.

\paragraph{Integration over the outer sample}\label{sec:outer-variance}
We now return to unconditional expectations, including the randomness in the outer input and permutation. The conditional quantities $L_\ell$ and $b_\ell$ discussed so far depend on $(\bsX,\pi)$. Even when meeting tails are geometric for each outer pair, their rates may approach one as the outer input varies. The next theorem therefore also requires integrability of the conditional bounds.

\begin{theorem}[Finite-variance unbiased estimates]\label{thm:finite-variance}
Assume that the coupling is faithful, every $\tau_{\ell}$ is almost surely finite, and \eqref{eq:mean-convergence} holds at every level for almost every outer pair. Suppose that, for a common $p\in(2,\infty]$, \eqref{eq:contrast-moment} holds and $b_\ell<\infty$ at every level for almost every outer pair. If
\begin{equation}\label{eq:outer-moment-condition}
 \EE\!\left[\sum_{\ell=1}^dL_\ell^2(1+2b_\ell)^2\right]<\infty,
\end{equation}
then the integrability condition \eqref{eq:L1-assumption} holds, every $A_\ell$ and $D_j$ has finite second moment, and the conclusions of \cref{thm:unbiased} apply.
\end{theorem}
\begin{proof}
Since $\psi_{\bsX}(\bsX)=0$, telescoping across levels gives
\[
 \psi_{\bsX}(\bsZ_{\ell,t})=\sum_{i=1}^{\ell}G_{i,t}.
\]
The difference between the main and auxiliary evaluations likewise satisfies
\[
    \psi_{\bsX}(\bsZ_{\ell,t})
    -\psi_{\bsX}(\widetilde{\bsZ}_{\ell,t-1})
    =\sum_{i=1}^{\ell}
      (G_{i,t}-\widetilde G_{i,t-1}).
\]
The triangle inequality and \cref{prop:adjacent-moments} therefore bound the conditional $L^2$ norm of the nonnegative expression inside the expectation in \eqref{eq:L1-assumption} by $\sum_{i=1}^{\ell}L_i(1+2b_i)$. Its unconditional second moment is consequently at most
\[
    \ell\,\EE\!\left[
        \sum_{i=1}^{\ell}L_i^2(1+2b_i)^2
    \right]<\infty.
\]
This proves \eqref{eq:L1-assumption} and the finite second moments of $A_\ell$ and its adjacent differences. The expectations then follow from \cref{thm:unbiased}.
\end{proof}

Under these assumptions, the conditional means of the contributions are determined by the target distributions.  The law of total variance gives
\begin{equation*}
 \VV[\widehat\phi_{j,N}]
 =\frac1N\left\{
   \VV[\EE[D_j\,|\,\bsX,\pi]]+
   \EE[\VV[D_j\,|\,\bsX,\pi]]\right\}.
\end{equation*}
The first term is fixed by the outer sampling scheme. Conditional simulation and cross-level coupling affect the second term, which is bounded using \eqref{eq:variance-bound} at the random level $\ell=\pi^{-1}(j)$.

\paragraph{Preservation of other coordinates}
We next consider an input that is independent of the remaining inputs, which may still be mutually dependent. If the shared updates preserve agreement in all other coordinates, the variation of the response in this input controls the second moment of its estimated contribution. Expectations in the following statement are unconditional; $b_\ell$ remains the conditional quantity defined in \eqref{eq:tail-sum}.

\begin{proposition}[Variation in one independent input]
\label{prop:coordinate-variation}
Suppose that $X_j$ is independent of $\bsX_{-\{j\}}$ and
$\omega_f=\sup f-\inf f<\infty$. Define
\[
 \delta_j=
 \sup_{\bsz_{-\{j\}}=\bsy_{-\{j\}}}
       |f(\bsz)-f(\bsy)|,
\]
where the supremum is taken over feasible input vectors. For almost every outer pair, suppose that the shared updates at the adjacent levels with $j=\pi(\ell)$ preserve agreement outside coordinate $j$ whenever the current feasible states agree there. The state at level zero remains $\bsx$. Assume that the lagged coupling is faithful and the meeting times are almost surely finite. Define $b_\ell$ using $p=\infty$ in \eqref{eq:tail-sum}, so that $\alpha=1/2$. If $\EE[(1+2b_{\pi^{-1}(j)})^2]<\infty$, then
\begin{equation}\label{eq:coordinate-variation-bound}
 \EE[D_j^2]
 \le \omega_f^2\delta_j^2
       \EE[(1+2b_{\pi^{-1}(j)})^2].
\end{equation}
If $\delta_j=0$, then $D_j=0$ almost surely, without requiring the integrability assumption on $b_{\pi^{-1}(j)}$.
\end{proposition}

\begin{proof}
Fix an outer pair and put $\ell=\pi^{-1}(j)$. Both adjacent main chains start at $\bsx$. Induction through the shared updates shows that their states agree outside $j$ at every time. The same argument applies to the indefinitely continued auxiliary chains. For $\ell=1$, all coordinates other than $j$ remain fixed. By \eqref{eq:factorization},
\[
 |G_{\ell,t}|\le\omega_f\delta_j,
 \qquad L_\ell\le\omega_f\delta_j.
\]
Apply \cref{prop:adjacent-moments} conditionally with $p=\infty$ and integrate over the outer pair to obtain \eqref{eq:coordinate-variation-bound}. If $\delta_j=0$, both $G_{\ell,t}$ and $\widetilde G_{\ell,t}$ vanish at every time. \eqref{eq:D-decomposition} then gives $D_j=0$, including the meeting correction, even when the adjacent meeting times differ.
\end{proof}

Under these preservation assumptions, an independent input absent from the response has an identically zero estimated contribution. When $\delta_j=0$ and the corrected level estimates have finite second moments, the matched independent-level construction in \cref{prop:independent-level-comparison} instead satisfies
\[
 \VV[D_j^{\mathrm{ind}}]
 =2\EE\!\left[
   \VV[A_{\pi^{-1}(j)}\,|\,\bsX,\pi]\right].
\]
Indeed, the adjacent corrected estimates coincide in the shared construction, so their conditional marginal laws agree. The displayed variance is positive whenever the conditional level estimate is nondegenerate with positive probability. More generally, for bounded functions $g$ and $h$, consider $f_\varepsilon(\bsx)=g(\bsx_{-\{j\}})+\varepsilon h(\bsx)$. With the input law, kernels, coupling, and averaging window fixed, $\delta_j=O(|\varepsilon|)$ and the response range remains bounded as $\varepsilon\to0$. Hence \eqref{eq:coordinate-variation-bound} gives $\VV[D_j^{(\varepsilon)}]=O(\varepsilon^2)$, provided that the stated meeting-tail integrability condition holds, where $D_j^{(\varepsilon)}$ denotes the contribution estimate for $f_\varepsilon$.

\paragraph{Standard errors}
Let $\bsD=(D_1,\ldots,D_d)^\top$, $\boldsymbol\phi=(\phi_1,\ldots,\phi_d)^\top$, and $\Sigma=\Cov(\boldsymbol D)$. For $N\ge2$, let $\widehat\Sigma_N$ be
the sample covariance matrix, with divisor $N-1$, of the independent completed vectors in \eqref{eq:outer-estimator}. Under \cref{thm:finite-variance}, $\Cov(\widehat{\boldsymbol\phi}_N)=\Sigma/N$, and the usual multivariate central limit theorem applies. The strong law also gives $\widehat\Sigma_N\to\Sigma$ almost surely \cite{V98}. Thus, $\sqrt{\widehat\Sigma_{N,jj}/N}$ estimates the standard error of $\widehat\phi_{j,N}$, and the full covariance matrix also gives standard errors for differences between contributions.

For normalized effects, set $\widehat{\tilde\phi}_{j,N}=\widehat\phi_{j,N}/\widehat\sigma_N^2$ when $\widehat\sigma_N^2\ne0$, and set it to zero otherwise. These estimates are consistent because $\widehat\phi_{j,N}\to\phi_j$ and $\widehat\sigma_N^2\to\sigma^2>0$ almost surely, but they need not be unbiased at finite $N$. For a nonzero denominator, the delta method \cite[Chapter~3]{V98} gives the estimated squared standard error
\begin{equation}\label{eq:normalized-se}
 \frac{1}{N(N-1)(\widehat\sigma_N^2)^2}
 \sum_{r=1}^N
 \left\{D_j^{(r)}-\widehat{\tilde\phi}_{j,N}A_d^{(r)}\right\}^2.
\end{equation}
This expression includes the covariance with the estimated total variance. The normalized estimates sum to one whenever the denominator is nonzero. \noeqref{eq:normalized-se}
\subsection{Computational cost}\label{sec:conditional-work}
At level $\ell$, \cref{alg:estimator} advances the main chain through time $\max(m,\tau_\ell)$ and the auxiliary chain through time $\tau_\ell-1$. The total number of marginal transitions at that level is therefore $\max(m,\tau_\ell)+\tau_\ell-1$.

Let $\mathcal C$ denote the work required to generate one complete contribution vector. Let $c_{\mathrm{out}}$ bound the expected one-off work conditional on $(\bsx,\pi)$, including generation of the outer pair, preparation of the update maps, evaluation of $f(\bsx)$, and final assembly of the contribution vector. Allocate the costs of shared random-block generation and bookkeeping to the marginal transitions, and suppose that a transition at level $\ell$, including any required response evaluation and meeting check, costs at most $c_\ell$ under this accounting. These nonnegative bounds $c_{\mathrm{out}}$ and $c_\ell$ may depend measurably on the outer pair. Since $\max(m,\tau_\ell)=m+(\tau_\ell-m)_+$, 
\begin{equation}\label{eq:expected-work}
 \EE[\mathcal C\,|\,\bsx,\pi]
 \le c_{\mathrm{out}}+
 \sum_{\ell=1}^d c_\ell
 \left\{m+\EE[\tau_\ell-1\,|\,\bsx,\pi]
           +\EE[(\tau_\ell-m)_+\,|\,\bsx,\pi]\right\},
\end{equation}
where $(a)_+=\max\{a,0\}$. For integer-valued meeting times,
\[
 \EE[\tau_\ell-1\,|\,\bsx,\pi]
   =\sum_{t=1}^{\infty}\PP(\tau_\ell>t\,|\,\bsx,\pi),\qquad
 \EE[(\tau_\ell-m)_+\,|\,\bsx,\pi]
   =\sum_{t=m}^{\infty}\PP(\tau_\ell>t\,|\,\bsx,\pi).
\]
Integrability of the right-hand side of \eqref{eq:expected-work} over $(\bsX,\pi)$ gives finite unconditional expected work. This is separate from the finite-variance requirement in \eqref{eq:outer-moment-condition}.

\paragraph{Efficiency and parameter selection}\label{sec:efficiency}\label{sec:pilot}
Under \cref{thm:finite-variance}, suppose also that
$0<\EE[\mathcal C]<\infty$. For a fixed number $N$ of independent completed runs,
\begin{equation*}
 \EE\!\left[\|\widehat{\boldsymbol\phi}_N-\boldsymbol\phi\|_2^2\right]
       =\frac{\tr\Sigma}{N},\qquad
 \EE[\text{total work}]=N\EE[\mathcal C].
\end{equation*}
Thus $\EE[\mathcal C]\,\tr\Sigma$ measures the product of mean squared error and expected work for the unnormalized contribution vector; smaller values indicate greater efficiency under this criterion. This product is standard in simulation efficiency \cite{GW92,JOA20}. The identity remains valid when the work and the output within a run are dependent, since the number of completed runs is fixed. The comparison depends on the chosen work measure: response evaluations, actual solver calls, and elapsed time can give different rankings.

An independent pilot may select a parameter configuration from a finite collection, all satisfying the preceding assumptions for the same input law and response. Production then uses fresh random streams and a sample count fixed before the production runs begin. Conditional on the pilot, the production vectors are independent and identically distributed with mean $\boldsymbol\phi$, so their average remains unbiased. 
The variance, standard error, and efficiency statements apply conditionally on the selected configuration and sample count.

\section{Numerical experiments}\label{sec:numerical}
We use three test cases to illustrate the computational efficiency of the proposed method: an Ising model with enumerated reference values, a hierarchical flood model requiring inverse conditioning, and a latent-Gaussian material model with a finite-element response. Dedicated exact conditional samplers provide verification and direct-generation benchmarks for these examples.

\subsection{Methods and experimental design}\label{sec:protocol}
\Cref{tab:methods} summarizes the methods compared in numerical experiments. Coupled unbiased MCMC (CU) refers to our proposed \cref{alg:estimator}. Independent-level unbiased MCMC (IU) uses the same initial states, marginal kernels, within-level lagged couplings, averaging window, and meeting correction as CU, but uses independent simulation streams across levels conditional on $(\bsX,\pi)$. Both reuse each level estimate in the adjacent contributions. Thus, their comparison isolates the effect of cross-level sharing, as in \cref{prop:independent-level-comparison}. Matched finite MCMC uses the CU main paths and averaging window without the meeting correction. For the flood and material models, we also use a longer finite-MCMC run with shared randomness and $(k,m)=(8,32)$, as well as exact conditional rejection sampling with shared (DS) or independent (DI) randomness across levels.

\begin{table}[t]
\caption{Comparison methods. The first three rows apply to all examples; the last three apply to the flood and material examples. The Ising example also uses conditional
pick-freeze and nested-variance comparators. Detailed explanations are in \cref{app:protocol-details}.}
\label{tab:methods}\centering\small
\setlength{\tabcolsep}{3pt}
\begin{tabular}{>{\raggedright\arraybackslash}p{.28\linewidth}>{\raggedright\arraybackslash}p{.30\linewidth}>{\raggedright\arraybackslash}p{.17\linewidth}>{\raggedright\arraybackslash}p{.15\linewidth}}
\toprule
Method & Conditional estimate & Cross-level randomness & Meeting correction\\
\midrule
CU & MCMC time average & Shared & Yes\\
IU & MCMC time average & Independent & Yes\\
Finite, matched & CU main-path time average & Shared & No\\
Finite, long & MCMC, $(k,m)=(8,32)$ & Shared & No\\
Direct, shared (DS) & Exact rejection draws & Shared & Not needed\\
Direct, independent (DI) & Exact rejection draws & Independent & Not needed\\
\bottomrule
\end{tabular}
\end{table}

Let $N$ be the number of completed contribution vectors per estimate and $R$ the number of independent replications. The Ising model uses $R=50$ and $N\in\{128, 512, 2048, 8192\}$, whereas the flood and material models use $R=30$ and $N=512$ for the common-sample-size comparison. Independent pilots select $(k,m)=(2,8)$ for the Ising model with dependent inputs ($\beta=0.35$ in \eqref{eq:ising-law}) and $(k,m)=(1,1)$ for the flood and material models. For the independent-input Ising case ($\beta=0$), we use direct refreshes with $k=m=1$. Production sample sizes are fixed in advance, and all prescribed contribution vectors are completed.

For a reference vector $\boldsymbol\phi^{\rm ref}$, the component-averaged root-mean-square error (RMSE) is
\begin{equation}\label{eq:empirical-rmse}
 \left\{\frac1{dR}\sum_{r=1}^R\sum_{j=1}^d
 (\widehat\phi_{j,N}^{(r)}-\phi_j^{\rm ref})^2\right\}^{1/2}.
\end{equation}
For the Ising model, the reference is the exact Shapley vector obtained by enumeration. For each flood or material setting, we construct a Monte Carlo reference by averaging 32,768 independent contribution vectors generated using exact conditional sampling. This reference is computed independently of the experimental runs but is used in common across all methods and sample sizes within the same setting.

Variance comparisons use the trace of the contribution-vector covariance matrix. We measure computational work by the number of model-response evaluations, including finite-element solves in the material model. For the flood model, we also compare the methods at approximately matched expected response budgets, with sample sizes chosen by independent cost pilots. These costs exclude pilot and reference calculations. Further details on sampling, standard errors, reference uncertainty, and cost accounting are provided in the supplementary material; see \cref{app:protocol-details,app:execution}.

\subsection{Ising model}\label{sec:ising}
We consider eight spins $\bsx\in\{\pm 1\}^8$ with distribution
\begin{equation}\label{eq:ising-law}
 \rho(\bsx)\ \propto\
 \exp\left\{\sum_{i=1}^8h_ix_i+\beta\sum_{i=1}^8x_ix_{i+1}\right\},
 \qquad x_9=x_1,\qquad h_i=-\frac{1}{4}+\frac{i-1}{14}.
\end{equation}
We examine the cases with $\beta=0$ (independent inputs) and $\beta=0.35$. The response is
\begin{equation*}
 f(\bsx)=\sum_{i=1}^8 0.7^{i-1}x_i
            +0.6x_1x_2+0.4x_3x_5+0.3x_2x_4x_6.
\end{equation*}
Enumeration over all 256 states gives the exact Shapley effects and permits exact outer sampling. At level $\ell$, the coordinates in $-U_\ell$ remain fixed at their outer values, and each sweep updates the free coordinates in $U_\ell$ in physical order $1,\ldots,8$ using Gibbs transitions \cite{RR04}. With cyclic subscripts $z_0=z_8$ and $z_9=z_1$, the conditional probability of setting spin $i$ to $+1$ is
\begin{equation}\label{eq:ising-conditional}
 p_i(\bsz_{-\{i\}})
 =\frac{1+\tanh\{h_i+\beta(z_{i-1}+z_{i+1})\}}2.
\end{equation}
Each update sets the spin to $+1$ if a uniform random variable is at most $p_i(\bsz_{-\{i\}})$, and to $-1$ otherwise. CU shares these uniforms across levels and replicas, while IU shares them only within each level. Each sweep preserves $\mu_\ell^{\bsx}$.

\begin{proposition}[Meeting for the binary sweep]\label{prop:ising-meeting}
For finite $h_i$ and $\beta$, the common-uniform coupling has a uniformly positive one-sweep meeting probability, and hence geometrically tailed meeting times. If $0<2\tanh|\beta|<1$, then
\begin{equation}\label{eq:ising-sharp}
 \PP(\tau_\ell>s+1\,|\,\bsx,\pi)
 \le\min\{1,\ell(2\tanh|\beta|)^s\},\qquad s\ge0.
\end{equation}
\end{proposition}
\begin{proof}
The minimum of \eqref{eq:ising-conditional} over all configurations is positive. If every uniform in a paired sweep is below the corresponding minimum, all free spins become $+1$ in both replicas, while the fixed spins already agree. This gives a uniformly positive meeting probability. Changing one neighboring spin changes the conditional probability by at most $\tanh|\beta|$. Under common-uniform updates, the disagreement probability equals the difference of the conditional probabilities. If all free-spin disagreement probabilities are at most $r$ before a sweep, each updated spin has disagreement probability at most $2r\tanh|\beta|$. Induction through the systematic scan gives this bound for every free spin after the sweep. Starting from $r=1$, iterating over $s$ sweeps, and applying a union bound over the $\ell$ free coordinates proves \eqref{eq:ising-sharp}; see \cite{DGJ08}.
\end{proof}

Coupling to a stationary chain gives marginal convergence. Since the response is bounded and meeting times have geometric tails, the required moment and expected-work conditions hold. For $\beta=0$ and $k=m=1$, \cref{cor:independent} recovers coordinate swapping samplewise.

\subsection{A hierarchical two-period flood model}\label{sec:flood}
As the second test case, we use the hydraulic response and five hydraulic marginal distributions in \cite[Section~6.1]{ICI21}, and specify a new two-period hierarchy for the flood loads. The treatment of uncertain load-distribution parameters is motivated by the engineering model in \cite[Section~6.3]{EPS24}.

\subsubsection{Physical inputs and response}
The nine inputs are
\[
 \bsX=(A_Q,B_Q,Q_1,Q_2,K_s,Z_v,Z_m,L,B_{\mathrm{width}}).
\]
For $\eta\in\{0.5,1.5\}$, the independent hyperparameters follow
\begin{equation*}
 A_Q\sim\operatorname{Unif}[1000-200\eta,1000+200\eta],\qquad
 B_Q\sim\operatorname{Unif}[550-150\eta,550+150\eta].
\end{equation*}
Conditional on $(A_Q,B_Q)=(a,b)$, the two loads $Q_1,Q_2$ are independent $\operatorname{Gumbel}(a,b)$ variables, each truncated to $[500,3000]$. Writing $G(t)=\exp\{-\exp(-t)\}$ and $g(t)=\exp\{-t-\exp(-t)\}$, their conditional density is
\begin{equation}\label{eq:flood-density}
 p(q\,|\, a,b)=
 \frac{g((q-a)/b)}{b\,Z(a,b)}\mathds{1}_{[500,3000]}(q),\qquad
 Z(a,b)=G((3000-a)/b)-G((500-a)/b).
\end{equation}
The normalizing factor $Z(a,b)$ depends on both hyperparameters and is included in every likelihood evaluation. The remaining inputs are mutually independent of the load hierarchy and of one another. They follow the hydraulic marginals
\[
\begin{aligned}
 K_s&\sim N(30,7^2)\text{ truncated to }[15,\infty),\\
 Z_v&\sim\operatorname{Tri}(49,50,51),\qquad
 Z_m\sim\operatorname{Tri}(54,55,56),\\
 L&\sim\operatorname{Tri}(4990,5000,5010),\qquad
 B_{\mathrm{width}}\sim\operatorname{Tri}(295,300,305),
\end{aligned}
\]
where $\operatorname{Tri}(a,c,b)$ denotes the triangular distribution with endpoints $a,b$ and mode $c$. The response is the maximum water level over the two periods:
\begin{equation}\label{eq:flood-response}
 f(\bsX)=Z_v+
 \left\{\frac{\max(Q_1,Q_2)}
 {B_{\mathrm{width}}K_s\sqrt{(Z_m-Z_v)/L}}\right\}^{3/5}.
\end{equation}
Exact outer sampling follows the generative hierarchy. The two values of $\eta$ define different input distributions and hence different Shapley effects.

When $Q_1$ is retained but the hyperparameters and $Q_2$ are released, the conditional law of $Q_2$ averages over the posterior of $(A_Q,B_Q)$ given $Q_1$. Thus the hyperparameters influence the Shapley effects even though they do not appear explicitly in \eqref{eq:flood-response}.

\subsubsection{Generative proposals and fixed-load likelihoods}
At level $\ell$, the independence proposal draws released hyperparameters from their priors, released loads from \eqref{eq:flood-density} given the proposed or retained hyperparameters, and released hydraulic inputs from their marginals. Retained coordinates remain fixed at their outer values.

Let $J_\ell$ be the indices of the retained loads. If $J_\ell=\emptyset$ or both hyperparameters are retained, the proposal is an exact conditional draw. Otherwise, the target-to-proposal ratio reduces, up to a constant, to
\begin{equation}\label{eq:flood-weight}
 w_\ell(\bsz)=\prod_{j\in J_\ell}p(q_j^{\bsx}\mid a,b),
\end{equation}
where $q_j^{\bsx}$ is the retained outer value of $Q_j$, and $(a,b)$ denotes the hyperparameter values in $\bsz$. An independence Metropolis--Hastings update accepts a proposal $\bsz'$ with probability $\min\{1,w_\ell(\bsz')/w_\ell(\bsz)\}$. The proposal and acceptance randomness are shared according to the CU and IU rules in \cref{sec:protocol}.

\begin{proposition}[Meeting for the hierarchical flood kernel]
\label{prop:flood-meeting}
For either specified value of $\eta$, the common-proposal coupling has a positive lower bound on its one-step meeting probability, uniformly over the outer input, permutation, level, and feasible pair of states. The corrected estimators have finite moments of every order and finite expected simulation work.
\end{proposition}
\begin{proof}
Write $[a_-,a_+]$ and $[b_-,b_+]$ for the prior ranges. Here $500<a_-\le a_+<3000$ and $b_->0$. Monotonicity of $G$ gives
\[
 Z(a,b)\ge Z_*:=G((3000-a_+)/b_+)-G((500-a_-)/b_+)>0.
\]
Since $\sup_t g(t)=1/e$, we obtain the finite envelope
\begin{equation}\label{eq:flood-envelope}
 0< p(q\,|\, a,b)\le
 M:=\frac{1}{e\,b_-Z_*},\qquad q\in[500,3000].
\end{equation}
Put $n_\ell=|J_\ell|\le2$. For every feasible current state, the Metropolis acceptance probability is at least $w_\ell(\bsz')/M^{n_\ell}$. A common uniform below this bound makes both replicas accept the same proposal. Since \eqref{eq:flood-density} has a positive minimum on the compact parameter and load ranges, this event has uniformly positive probability. The exact-refresh cases meet in one paired update. Thus the meeting times have uniformly geometric tails. The hydraulic inputs satisfy the deterministic response bound
\[
 49\le f(\bsX)\le 51+
 \left\{\frac{3000}{295\cdot15\sqrt{3/5010}}\right\}^{3/5}.
\]
The same coupling mechanism gives convergence to the invariant conditional distributions. Bounded responses and geometric meeting tails ensure finite moments of every order. As each transition has finite expected work, the conclusions follow from \cref{thm:unbiased,thm:finite-variance} and \eqref{eq:expected-work}.
\end{proof}

\begin{corollary}[Coupling of the hydraulic inputs]
\label{cor:flood-coordinate}
For each of $K_s,Z_v,Z_m,L,B_{\mathrm{width}}$, the shared updates satisfy the coordinate-preservation condition of \cref{prop:coordinate-variation}, and \eqref{eq:coordinate-variation-bound} holds. For the input $X_j=L$,
\begin{equation}\label{eq:flood-length-variation}
 \delta_j\le
 \left(\frac{3000}{295\cdot15\sqrt3}\right)^{3/5}
 \left(5010^{3/10}-4990^{3/10}\right)<0.00880.
\end{equation}
\end{corollary}
\begin{proof}
Releasing a hydraulic coordinate does not change the retained-load likelihood in \eqref{eq:flood-weight}. If adjacent states agree outside this coordinate, shared coordinate-indexed proposals and the common acceptance uniform preserve their agreement. The geometric meeting bound in \cref{prop:flood-meeting} ensures the required integrability. Finally, the dependence on $L$ in \eqref{eq:flood-response} is through $L^{3/10}$, which gives \eqref{eq:flood-length-variation}.
\end{proof}

The finite envelope $M$ in \eqref{eq:flood-envelope} also gives an exact conditional rejection sampler: accept the generative proposal with probability $w_\ell(\bsz')/M^{n_\ell}$, where $n_\ell=|J_\ell|$, and use direct refresh in the cases specified above. This sampler provides independent Monte Carlo reference estimates and the direct-generation comparisons.

\subsection{A two-phase material model}\label{sec:materials}
As our final test case, we allocate the variance of a specimen's apparent conductivity among nine physical cells. The binary material is obtained by thresholding a Gaussian field, following \cite{RT95}.

\subsubsection{Material configuration and response}
Partition the unit square $\mathcal D=(0,1)^2$ into $3\times3$ cells with centers $\boldsymbol r_a$, $a=1,\ldots,9$, indexed row by row. For $r_c\in \{0.12, 0.30\}$, let
\begin{equation*}
 \boldsymbol W\sim N_9(0,\mathsf K),\qquad
 \mathsf K_{ab}=\exp\{-\|\boldsymbol r_a-\boldsymbol r_b\|_2/r_c\},
 \qquad X_a=\mathds{1}\{W_a>0\}.
\end{equation*}
Each cell is one binary input, so $d=9$. Define $\mathcal B(\boldsymbol w)=(\mathds{1}\{w_a>0\})_{a=1}^9$; then $\bsX=\mathcal B(\boldsymbol W)$. A Gaussian draw followed by $\mathcal B$ generates independent outer samples.

The cell conductivities are $\kappa_0=1$ and $\kappa_1=10$, with $\kappa(\boldsymbol r;\bsX)=\kappa_0+(\kappa_1-\kappa_0)X_a$ on cell $a$. The potential is zero on the left boundary and one on the right boundary, with zero normal flux on the horizontal boundaries. We use its conforming bilinear finite-element solution $u_h$ on a uniform mesh with width $h=1/6$ \cite{C02}, defined by
\begin{equation*}
 \int_{\mathcal D}\kappa(\boldsymbol r;\bsX)
     \nabla u_h\cdot\nabla v_h\,\mathrm{d}\boldsymbol r=0
\end{equation*}
for all finite-element test functions $v_h$ vanishing on the two vertical boundaries. Each material cell contains $2\times2$ elements. Element integration is exact for the constant cell conductivities. The response is the discrete Dirichlet energy
\begin{equation*}
 f(\bsX)=\int_{\mathcal D}\kappa(\boldsymbol r;\bsX)
                       |\nabla u_h|^2\,\mathrm{d}\boldsymbol r,
\end{equation*}
which represents the finite-element apparent conductivity. The variational principle gives $1\le f(\bsX)\le10$. All methods use this same fixed discretized response.

The zero threshold gives phase probability $1/2$ at every cell, while the phase fraction of an individual specimen varies. This is a reduced two-dimensional member of the thresholded-Gaussian model family, with the geometry and boundary conditions specified here; \cite{RT95} also studies three-dimensional specimens with realization-wise volume-fraction control.

\subsubsection{Conditional simulation in Gaussian coordinates}
Given the outer phase vector $\bsx$ and permutation, the cells in $-U_\ell$ retain their phases. The latent conditional density is
\begin{equation}\label{eq:material-conditional}
 \nu_\ell^{\bsx}(\boldsymbol w)\ \propto\
 \varphi_{\mathsf K}(\boldsymbol w)
 \prod_{a\in -U_\ell}
 \mathds{1}\{\mathds{1}\{w_a>0\}=x_a\},
\end{equation}
where $\varphi_{\mathsf K}$ is the $N_9(0,\mathsf K)$ density. Its image under $\mathcal B$ is $\mu_\ell^{\bsx}$. Every latent coordinate remains in the conditional state; retained phases impose sign constraints and their latent magnitudes are integrated over.

At level $\ell$, we use an untilted separation-of-variables (SOV) independence proposal $q_\ell$, which visits constrained coordinates first in physical cell order, followed by unconstrained coordinates. After sign changes, a Cholesky factorization expresses the constraints as sequential lower bounds $a_{\ell,j}$ on standard normal innovations. The proposal samples constrained innovations from the corresponding truncated normal laws and unconstrained innovations from standard normal laws. Its unnormalized target-to-proposal weight is
\begin{equation}\label{eq:material-weight}
\begin{aligned}
 w_\ell(\boldsymbol w) =\frac{\varphi_{\mathsf K}(\boldsymbol w)
   \prod_{a\in -U_\ell}\mathds{1}\{\mathds{1}\{w_a>0\}=x_a\}}
       {q_\ell(\boldsymbol w)} =\prod_{j=1}^{d-\ell}\Phi\{-a_{\ell,j}(\boldsymbol w)\},
 \qquad 0<w_\ell\le1,
\end{aligned}
\end{equation}
where $\Phi$ is the standard normal CDF and the empty product is one. See \cite{B17} for SOV proposals and minimax tilting.

An independence Metropolis–Hastings update accepts a proposal $\boldsymbol w'\sim q_\ell$ with probability $\min\{1,w_\ell(\boldsymbol w')/w_\ell(\boldsymbol w)\}$. Cell-indexed proposal random numbers and the acceptance uniform follow the CU and IU sharing rules in \cref{sec:protocol}. Every level starts from the same feasible latent vector $w_a=2x_a-1$. The response evaluations in \eqref{eq:A} use $\bsZ_{\ell,t}=\mathcal B(\boldsymbol W_{\ell,t})$ and $\widetilde{\bsZ}_{\ell,t}=\mathcal B(\widetilde{\boldsymbol W}_{\ell,t})$. Meeting is checked on complete latent states; equality of phase patterns alone is insufficient:
\[
 \tau_\ell=\inf\{t\ge1:
   \boldsymbol W_{\ell,t}=\widetilde{\boldsymbol W}_{\ell,t-1}\}.
\]
The augmented-state construction follows \cref{rem:augmented-state}.

\begin{proposition}[Meeting for the material kernel]\label{prop:material-meeting}
For a fixed outer phase configuration and permutation, let $p_\ell^{\bsx}$ be the Gaussian probability of the sign restrictions in \eqref{eq:material-conditional}. The common-proposal coupling has one-step meeting probability at least $p_\ell^{\bsx}>0$ from every pair of feasible latent states. For this fixed nine-cell grid and either specified correlation length, the corrected estimators have finite moments of every order and finite expected simulation work.
\end{proposition}

\begin{proof}
By \eqref{eq:material-weight}, both acceptance probabilities are at least $w_\ell(\boldsymbol w')$. A common uniform below this value makes both replicas accept the same candidate. Its probability is
\[
 \int q_\ell(\boldsymbol w')w_\ell(\boldsymbol w')\,\mathrm{d}\boldsymbol w'
 =p_\ell^{\bsx}>0.
\]
Positive definiteness of $\mathsf K$ gives positive probability to each orthant. There are finitely many outer phase patterns, cell subsets, and permutations. Their meeting probabilities consequently have a positive minimum for either fixed correlation length. This yields uniformly geometric meeting tails. Coupling to a stationary copy gives convergence of expectations of the bounded function $\psi_{\bsx}\circ\mathcal B$. The response bound makes every correction term bounded, and a geometrically tailed number of terms has moments of every order. Each fixed-dimensional transition has finite work. Apply \cref{thm:unbiased,thm:finite-variance} and \eqref{eq:expected-work}.
\end{proof}

The same proposal gives an exact conditional rejection sampler by accepting $\boldsymbol w'\sim q_\ell$ with probability $w_\ell(\boldsymbol w')$, without requiring $p_\ell^{\bsx}$. We use this untilted SOV sampler as the exact-conditional reference and direct-generation benchmark. It is not the optimized minimax-tilted sampler of \cite{B17}, so the comparison does not establish superiority over the best available direct conditional methods.

\subsection{Numerical results}\label{sec:new-results}

\subsubsection{Ising results}\label{sec:ising-values}
For $\beta=0.35$, the CU estimates agree with the exact values: all eight component means lie within $1.6$ Monte Carlo standard errors of their enumerated counterparts (see \cref{tab:ising-contributions} in the supplementary material). The exact output variance is $\sigma^2=3.3714428978$. Finite-state calculations give an expected sum of $3.2778978623$ for matched finite MCMC with $(k,m)=(2,8)$, with root-mean-square component bias $0.0132505$. The corresponding empirical sums are $3.3698$ for CU and $3.2762$ for matched finite MCMC. Hence the correction removes a systematic fixed-window error that does not vanish as $N$ increases.

\subsubsection{Ising accuracy and cost}\label{sec:ising-performance}
\begin{figure}[t]
\centering
\includegraphics[width=.6\linewidth]{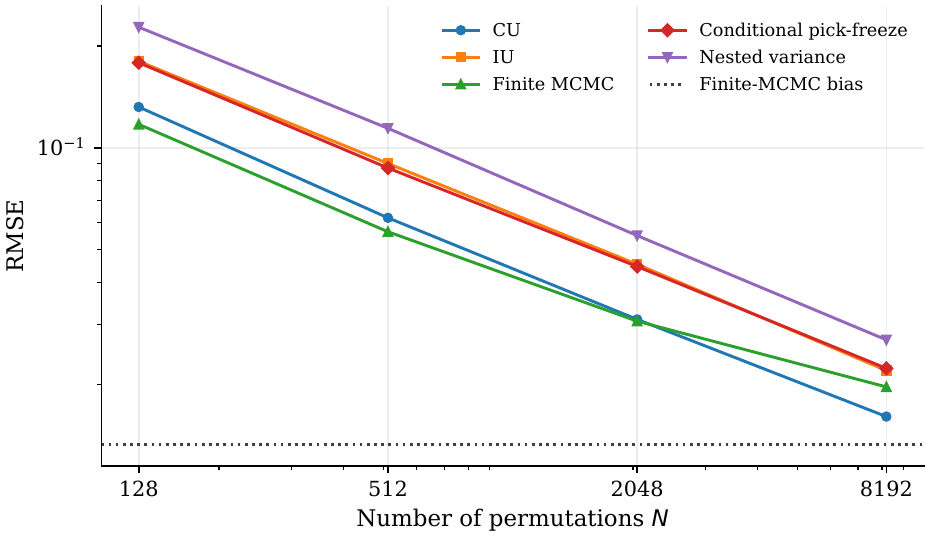}
\caption{Dependent Ising inputs, $\beta=0.35$, with $R=50$ independent experiment replications. CU and IU use $(k,m)=(2,8)$. The horizontal line is the root-mean-square bias of the finite-MCMC estimator, computed by finite-state transition calculations. The reference values are enumerated.}\label{fig:ising-scaling}
\end{figure}

\Cref{fig:ising-scaling} shows the RMSE against the exact reference for $\beta=0.35$. The fitted CU slope is $-0.506$, consistent with $N^{-1/2}$ convergence. At $N=8192$, the RMSEs of CU and matched finite MCMC are $0.01606$ and $0.01965$, respectively; at smaller sample sizes the finite estimator can have lower observed RMSE because of its smaller variance.

The IU/CU covariance-trace ratios are $3.19$ with bootstrap interval $[2.31,4.40]$ for $\beta=0$, and $1.88$ with interval $[1.50,2.36]$ for $\beta=0.35$. For the latter case, the conditional pick-freeze estimator is more efficient than CU in squared RMSE times response evaluations (see \cref{tab:ising-results} in the supplementary material). Thus MCMC debiasing is not necessarily preferable when efficient exact conditional draws are available.

\subsubsection{Flood results}\label{sec:flood-results}
For $\eta=0.5$ and $1.5$, the largest CU-estimated Shapley effect is that of $Z_v$, about $30.3\%$ and $31.8\%$ of the estimated output variance, respectively. The two loads and $K_s$ each contribute approximately $21\%$--$23\%$; the contribution of $L$ is near zero, consistent with \cref{cor:flood-coordinate}. Full componentwise estimates are given in \cref{tab:flood-contributions} of the supplementary material.

The hyperparameter effects are unresolved at $\eta=0.5$. For $\eta=1.5$, the normalized CU estimates for $A_Q$ and $B_Q$ are $0.908\%$ and $0.938\%$, with standard errors of $0.321$ and $0.353$ percentage points. \Cref{tab:flood-parents} shows that the matched finite estimates are negative, whereas the paired meeting corrections are positive by $3.5$--$6.9$ standard errors. The data do not establish an ordering between the two effects.

\begin{table}[t]
\caption{Meeting corrections for the flood hyperparameters. Both columns
are in squared response units, with one Monte Carlo standard error in
parentheses, using the same $30\times512$ vectors as
\cref{tab:flood-contributions}. The paired correction is CU minus matched
finite MCMC on the same simulated paths.}
\label{tab:flood-parents}\centering\small
\begin{tabular}{lcrr}
\toprule
$\eta$ & Input & Matched finite mean (SE) & Paired correction (SE)\\
\midrule
0.5 & $A_Q$ & $-0.00305\ (0.00070)$ & $0.00414\ (0.00094)$\\
0.5 & $B_Q$ & $-0.00255\ (0.00071)$ & $0.00370\ (0.00105)$\\
1.5 & $A_Q$ & $-0.00632\ (0.00136)$ & $0.01287\ (0.00186)$\\
1.5 & $B_Q$ & $-0.00459\ (0.00162)$ & $0.01136\ (0.00189)$\\
\bottomrule
\end{tabular}
\end{table}

\subsubsection{Material results}\label{sec:material-results}
For both $r_c=0.12$ and $0.30$, the central cell has the largest CU-estimated Shapley effect: $14.87\%$ and $15.02\%$ of the estimated output variance, with standard errors $0.48$ and $0.55$ percentage points, respectively. The estimated output variance increases from $3.3990$ to $5.5432$ as $r_c$ increases. These comparisons concern different input laws on the same nine-cell grid, not convergence in the number of cells or the finite-element resolution. Complete componentwise results are in \cref{tab:material-contributions} of the supplementary material.

\subsubsection{Variance and cost comparisons}\label{sec:joint-results}
\Cref{tab:new-efficiency} compares CU and IU in the flood and material models. The estimated CU covariance trace is approximately $60\%$--$78\%$ lower than that of IU, with all four pointwise bootstrap intervals favoring CU. For exact conditional sampling, the flood DI/DS variance ratios are $3.72$ and $3.05$ for $\eta=0.5$ and $1.5$, respectively.

\begin{table}[t]
\caption{CU--IU variance ratios for the four flood and material conditions.
$V$ is the mean within-replication sample covariance trace of the
$N=512$ contribution vectors. Brackets give pointwise 95\%
replication-bootstrap intervals. Ratios above one favor CU.}
\label{tab:new-efficiency}\centering\small
\setlength{\tabcolsep}{4pt}
\begin{tabular}{lc}
\toprule
Condition & $V_{\rm IU}/V_{\rm CU}$\\
\midrule
Flood, $\eta=0.5$ & $4.53\ [4.32,4.76]$\\
Flood, $\eta=1.5$ & $4.03\ [3.72,4.36]$\\
Material, $r_c=0.12$ & $3.71\ [3.57,3.86]$\\
Material, $r_c=0.30$ & $2.50\ [2.41,2.60]$\\
\bottomrule
\end{tabular}
\end{table}

At pilot-matched expected response-evaluation budgets, \cref{tab:budget-flood} shows nearly identical observed RMSE for CU and matched finite MCMC when $\eta=0.5$, and lower RMSE for matched finite MCMC when $\eta=1.5$. The longer finite window allows only $23$ or $24$ contribution vectors and performs worse at these budgets.

\begin{table}[t]
\caption{Flood accuracy at pilot-matched expected budgets. Expected response budgets are 5,156, 5,088, respectively. Counts $N$ are fixed by independent cost pilots. Cost is the attained mean response count per complete estimate; RMSE uses \eqref{eq:empirical-rmse}, with $R=30$.}
\label{tab:budget-flood}\centering\small
\begin{tabular}{lrrr rrr}
\toprule
& \multicolumn{3}{c}{Flood, $\eta=0.5$} & \multicolumn{3}{c}{Flood, $\eta=1.5$}\\
\cmidrule(lr){2-4}\cmidrule(lr){5-7}
Method & $N$ & Cost & RMSE & $N$ & Cost & RMSE\\
\midrule
CU & 512 & 5,109.1 & 0.01765 & 512 & 5,044.2 & 0.02144\\
IU & 513 & 5,126.5 & 0.03646 & 516 & 5,100.4 & 0.03488\\
Finite, matched & 523 & 5,150.3 & 0.01762 & 533 & 5,070.6 & 0.01975\\
Finite, long & 23 & 5,119.3 & 0.04619 & 24 & 5,164.4 & 0.04820\\
Direct, shared & 516 & 5,160.0 & 0.01861 & 509 & 5,090.0 & 0.02193\\
Direct, independent & 516 & 5,160.0 & 0.03385 & 509 & 5,090.0 & 0.03551\\
\bottomrule
\end{tabular}
\end{table}

At common sample sizes, the longer finite window has lower observed RMSE than CU at $N=512$ in all four flood and material settings, but requires more work per vector (see \cref{fig:new-accuracy} in the supplementary material). Their Monte Carlo reference noise floors are $0.002779$, $0.003072$, $0.01890$, and $0.02569$, respectively, contributing to the reported RMSE as described in \eqref{eq:reference-error}. Additional meeting and identity diagnostics are provided in \cref{app:experiment-diagnostics}.

\section{Concluding remarks}\label{sec:conclusion}

We have developed a permutation-based estimator of unnormalized Shapley effects by coupling conditional MCMC simulations across levels and using lagged replicas to remove finite-time bias. The method requires independent samples from the joint input distribution and invariant kernels for the conditional laws, but not exact conditional sampling. Under suitable assumptions, the resulting contribution vectors are unbiased, have finite variance and finite expected work, and sum samplewise to an unbiased estimator of the output variance.

The variance reduction from cross-level coupling is governed by the covariances between adjacent corrected level estimates. For an independent input whose shared updates preserve agreement in all other coordinates, we also established a second-moment bound in terms of its response variation. In particular, an input absent from the response has an identically zero estimated contribution under these conditions, even after meeting correction.

The numerical experiments illustrate the effects of both coupling mechanisms. Sharing randomness across levels reduced the covariance trace relative to independent-level simulation, while meeting correction eliminated finite-window bias. At the tested flood response budgets, matched short finite-MCMC runs achieved comparable or slightly lower observed RMSE than CU, illustrating the bias--variance tradeoff.

The hierarchical flood and latent-Gaussian material models illustrate conditional sampling problems that do not follow the natural generative order. Exact conditional samplers were available for validation, but are not required by our method. Our comparisons are limited to the implemented samplers and fixed model dimensions. Future work includes more challenging conditional models, comparisons with optimized direct samplers, and dimension-dependent bounds on meeting times and adjacent contributions.

\section*{Declaration of AI use}
The authors used ChatGPT Pro (OpenAI) to assist with language editing and proofreading and to obtain feedback on the manuscript. The second author acknowledges OpenAI for providing access to ChatGPT through the ChatGPT for Academic Researchers program to assist with the final review of the manuscript. The authors assume responsibility for all content. 

\FloatBarrier
\Needspace{8\baselineskip}


\clearpage
\setcounter{page}{1}
\setcounter{section}{0}
\setcounter{subsection}{0}
\setcounter{subsubsection}{0}
\setcounter{equation}{0}
\setcounter{theorem}{0}
\setcounter{figure}{0}
\setcounter{table}{0}
\setcounter{algorithm}{0}
\renewcommand{\siampretitle}{Supplementary Materials: }
\renewcommand{\siamprelabel}{SM}
\makeatletter
\let\maketitle\arxivmaketitle
\let\@maketitle\arxivinternalmaketitle
\let\thanks\arxivthanks
\makeatother
\flushbottom
\headers{Coupled MCMC for Shapley effects}{M. Hirono and T. Goda}

\title{Coupled unbiased MCMC for estimating Shapley effects under dependent inputs}

\author{Masataka Hirono\thanks{Graduate School of Engineering, The University of Tokyo, 7-3-1 Hongo, Bunkyo-ku, Tokyo 113-8656, Japan  (\email{mskhirono@g.ecc.u-tokyo.ac.jp}).}
\and Takashi Goda\thanks{Graduate School of Information Science and Technology, The University of Tokyo, 7-3-1 Hongo, Bunkyo-ku, Tokyo 113-8656, Japan (\email{takashigoda@g.ecc.u-tokyo.ac.jp}).}
}

\maketitle

\section{Variance comparison after correction}
\label{app:correction-comparison}
A variance reduction of the uncorrected time averages need not persist when the meeting corrections are added. We give a sufficient condition for this reduction to be preserved, using the decomposition into a time average and a meeting correction discussed in \cite[Section~3.1]{SM-JOA20}.

Consider the matched CU and IU constructions defined in
\cref{sec:protocol}. They use the same initial states, marginal kernels, within-level lagged couplings, and averaging window $1\le k\le m$. Conditional on the outer pair $(\bsX,\pi)$, they differ only in the dependence of the simulation randomness across levels. Put $n=m-k+1$. For $a\in\{\mathrm{CU},\mathrm{IU}\}$, let $\boldsymbol B^a$ denote the uncorrected contribution vector and $\boldsymbol D^a$ the corrected vector, so that
\[
 B_{\pi(\ell)}^a=\frac1n\sum_{t=k}^m G_{\ell,t}^a,
 \qquad
    D_{\pi(\ell)}^a=B_{\pi(\ell)}^a+R_\ell^a,
    \qquad \ell=1,\ldots,d,
\]
where $R_\ell^a$ is the correction in \eqref{eq:D-decomposition}. Let $L_\ell^a$ and $b_\ell^a$ be the conditional quantities in \eqref{eq:contrast-moment} and \eqref{eq:tail-sum} for construction $a$. Define
\[
 W_a=\EE\!\left[
      \tr\Cov(\boldsymbol B^a\,|\,\bsX,\pi)\right],
 \qquad
 \varepsilon_a=
 2\left\{\EE\!\left[\sum_{\ell=1}^d(L_\ell^a b_\ell^a)^2\right]\right\}^{1/2}.
\]
Thus, $W_a$ is the conditional simulation variance of the uncorrected vector, summed over inputs and averaged over the outer pair. It excludes the variability of the conditional mean.

\begin{proposition}[Preservation of variance reduction]
\label{prop:correction-comparison}
Suppose that both matched constructions satisfy the assumptions of \cref{thm:finite-variance}. Then, $W_a$ and $\varepsilon_a$ are finite, and 
\begin{equation}\label{eq:conditional-variance-perturbation}
 \left|
 \left\{\EE\!\left[
   \tr\Cov(\boldsymbol D^a\,|\,\bsX,\pi)\right]\right\}^{1/2}
 -\sqrt{W_a}\right|\le\varepsilon_a,
 \qquad a\in\{\mathrm{CU},\mathrm{IU}\}.
\end{equation}
Consequently, if
\begin{equation}\label{eq:correction-gap-condition}
 \sqrt{W_{\mathrm{IU}}}-\sqrt{W_{\mathrm{CU}}}
 >\varepsilon_{\mathrm{IU}}+\varepsilon_{\mathrm{CU}},
\end{equation}
then
$\tr\Cov(\boldsymbol D^{\mathrm{CU}})
 <\tr\Cov(\boldsymbol D^{\mathrm{IU}})$.
\end{proposition}

\begin{proof}
The moment assumptions ensure that both $\boldsymbol B^a$ and $\boldsymbol D^a$ have finite second moments and that $\varepsilon_a<\infty$. By \eqref{eq:D-decomposition} and \cref{prop:adjacent-moments},
\[
 \EE\!\left[\|\boldsymbol D^a-\boldsymbol B^a\|_2^2\right]
 =\EE\!\left[\sum_{\ell=1}^d(R_\ell^a)^2\right]
 \le 4\EE\!\left[\sum_{\ell=1}^d(L_\ell^a b_\ell^a)^2\right]
 = \varepsilon_a^2,
\]
where $\|\cdot\|_2$ denotes the Euclidean norm. For any square-integrable random vector $\boldsymbol Y$,
\[
    \EE\!\left[
        \|\boldsymbol Y-\EE[\boldsymbol Y\,|\,\bsX,\pi]\|_2^2
    \right]
    =\EE\!\left[\tr\Cov(\boldsymbol Y\,|\,\bsX,\pi)\right]
    \le\EE\!\left[\|\boldsymbol Y\|_2^2\right].
\]
Applying the reverse triangle inequality in $L^2$ to the conditionally centered vectors $\boldsymbol D^a$ and $\boldsymbol B^a$, and then applying this contraction to their difference, proves \eqref{eq:conditional-variance-perturbation}.

Under \eqref{eq:correction-gap-condition}, the upper bound for CU in \eqref{eq:conditional-variance-perturbation} is strictly below the lower bound for IU, which is then positive. Hence the expected conditional covariance trace of $\boldsymbol D^{\mathrm{CU}}$ is strictly smaller than that of $\boldsymbol D^{\mathrm{IU}}$. Moreover, \cref{thm:unbiased} gives, for both constructions,
\[
    \EE[D_{\pi(\ell)}^a\,|\,\bsX,\pi]
    =\mu_\ell^{\bsX}(\psi_{\bsX})
     -\mu_{\ell-1}^{\bsX}(\psi_{\bsX}).
\]
Their conditional mean vectors therefore agree. By the law of total covariance, the variability of this common conditional mean contributes the same term to both unconditional covariance traces. Adding that term proves the stated comparison.
\end{proof}

\section{Additional experimental details}\label{app:experiment-details}

\subsection{Sampling, comparators, and inference}\label{app:protocol-details}
\paragraph{Sampling and pairing}
Within each experiment replication, methods share outer inputs and
permutations at matching vector indices, except for the independent
anchors of the Ising nested-variance comparator. Internal streams are
independent between methods, except that matched finite MCMC uses the
CU main paths and averaging window. 
For DS, proposal random numbers and acceptance uniforms are shared across levels by rejection-attempt index. Each level retains its first accepted proposal.
Smaller Ising samples are prefixes
of the $N=8192$ run; flood and material curves use prefixes
$N\in\{8,32,128,512\}$. The flood expected-budget comparisons use
their separately prescribed prefixes of the completed production vectors.
Curve points within a replication are therefore dependent. Production
streams are independent of all pilots and reference calculations.

\paragraph{Ising comparators and uncertainty}
The means in \cref{tab:ising-contributions} average the $R=50$
independent complete estimates at $N=8192$. Their standard errors
are the sample standard deviations of these estimates divided by
$\sqrt R$. For the Sum column, the inputs are summed within each
replication before calculating the standard error.

\begin{table}[htbp]
\caption{Exact and estimated unnormalized Shapley contributions for dependent Ising inputs,
$\beta=0.35$, with standard errors in parentheses below the estimates.
CU and matched finite MCMC use $(k,m)=(2,8)$, $N=8192$, and $R=50$.}
\label{tab:ising-contributions}\centering\small
\setlength{\tabcolsep}{3pt}
\begin{tabular}{@{}l*{9}{c}@{}}
\toprule
Method & $X_1$ & $X_2$ & $X_3$ & $X_4$ & $X_5$ & $X_6$ & $X_7$ & $X_8$ & Sum\\
\midrule
Exact & $1.1269$ & $0.9702$ & $0.5913$ & $0.2673$ & $0.1570$ & $0.1051$ & $0.0486$ & $0.1048$ & $3.3714$\\
\midrule
CU & $1.1254$ & $0.9681$ & $0.5927$ & $0.2689$ & $0.1556$ & $0.1070$ & $0.0477$ & $0.1044$ & $3.3698$\\
 & $(0.0027)$ & $(0.0038)$ & $(0.0029)$ & $(0.0019)$ & $(0.0014)$ & $(0.0012)$ & $(0.0010)$ & $(0.0019)$ & $(0.0048)$\\
\midrule
Finite & $1.1076$ & $0.9449$ & $0.5742$ & $0.2563$ & $0.1494$ & $0.1022$ & $0.0437$ & $0.0978$ & $3.2762$\\
 & $(0.0024)$ & $(0.0035)$ & $(0.0023)$ & $(0.0016)$ & $(0.0012)$ & $(0.0008)$ & $(0.0008)$ & $(0.0016)$ & $(0.0042)$\\
\bottomrule
\end{tabular}
\end{table}
The conditional pick-freeze comparator adapts the construction in
\cite[Section~5.2]{SM-G21} by coupling the exact conditional draws through
common coordinate-indexed random variables. Following \cite{SM-SNS16}, the
nested-variance comparator uses one independent outer anchor and three
conditionally independent responses per nonterminal prefix. Adjacent
differences reuse these variance estimates. A sample variance of $N$
independent joint responses supplies the terminal component, shared
within the replication. The response count is $N\{1+3(d-1)\}$, and
reported errors include the terminal uncertainty. Ising efficiency
ratios use the covariance of the $R$ complete estimates at common $N$
and their mean total response cost. Percentile intervals use 40,000
joint bootstrap resamples of replications. The reported CU and finite-MCMC
coverage uses approximate pointwise 95\% intervals centered at each
replication's
estimate, with standard errors from its $N$ completed vectors and
Student $t$ critical values with $N-1$ degrees of freedom.

\paragraph{Flood and material inference}
We apply the standard-error formulas in \cref{sec:variance} to the pooled
$30\times512$ independent vectors, retaining component covariances for
totals. The paired-correction standard errors in \cref{tab:flood-parents}
are calculated from the vectorwise differences between CU and matched
finite MCMC, retaining their within-pair dependence. Normalized standard
errors use \eqref{eq:normalized-se}.
For each method, the estimated single-vector covariance trace is the
average of the $R$ within-replication sample covariance traces at $N=512$.
We denote this estimate by $V$ in \cref{tab:new-efficiency}.
Variance-ratio intervals use 2,000
percentile-bootstrap resamples of the 30 entire replications, jointly
across methods.

\paragraph{Reference uncertainty}
The flood and material reference estimates are unbiased and independent
of the production estimates. Averaging over both calculations gives
\begin{equation}\label{eq:reference-error}
 \EE[(\widehat\phi_{j,N}-\phi_j^{\rm ref})^2]
 =\EE[(\widehat\phi_{j,N}-\phi_j)^2]+\VV[\phi_j^{\rm ref}].
\end{equation}
The estimated reference noise floor is
$\{\tr\Cov(\boldsymbol\phi^{\rm ref})/d\}^{1/2}$. The covariance of the
reference mean is estimated by dividing the sample covariance of the
32,768 independent reference vectors by 32,768.

\paragraph{Response accounting}
For flood and material, each production vector has its own cache,
keyed by the full flood state
or material phase pattern; caches are not shared between vectors or
methods. The outer response is counted once per vector. Finite-MCMC
costs count only the additional responses required in the averaging
window. Proposal counts and elapsed times are also recorded.

\paragraph{Pilots}
The candidate windows are $(1,1)$, $(1,4)$, and $(2,8)$, scored by
estimated covariance trace times mean response count. The dependent
Ising pilot uses 20 replications of 4096 completed vectors per candidate
and selects the smallest mean score; see \cref{tab:ising-pilot}.
Each flood and material pilot uses 128 completed CU vectors per
candidate. For the reported flood budget comparisons, a separate
128-vector pilot per method estimates the mean response costs
$\widehat c_a$. It sets the common expected budget
$B=512\widehat c_{\rm CU}$ and fixes
$N_a=\operatorname{round}(B/\widehat c_a)$ for method $a$.
The attained mean production costs are reported alongside the errors in
\cref{tab:budget-flood}.

\begin{table}[htbp]
\caption{Independent-pilot scores for dependent Ising inputs. Scores estimate single-vector trace variance times mean response-evaluation count. Parentheses contain one Monte Carlo standard error across the 20 pilot replications. Each replication contains 4096 completed vectors.}\label{tab:ising-pilot}
\centering\small
\begin{tabular}{cr}
\toprule
$(k,m)$ & Score (standard error)\\
\midrule
$(1,1)$ & $1092.41\ (13.35)$\\
$(1,4)$ & $1233.82\ (22.94)$\\
$(2,8)$ & $905.92\ (8.38)$\\
\bottomrule
\end{tabular}
\end{table}

\subsection{Execution and accounting}\label{app:execution}
For flood and material, computations used Python 3.11.17, NumPy 1.26.4,
SciPy 1.17.1, Windows 10, seed 20261020, and four worker processes. The complete invocation
took approximately 65.9 minutes. Recorded per-vector times include
conditional-kernel preparation but exclude model-wide fixed-matrix setup,
pilots, reference calculation, and diagnostics. Concurrent execution
affects the recorded timings, so they are not used to rank serial runtime.

For the flood expected-budget allocations in \cref{tab:budget-flood},
the implemented minimum of two vectors and cap of 2,048 are inactive
in both conditions.

Material direct generation uses blocks of 32 proposals. All generated
candidates are counted, including unused candidates after an acceptance.
The implementation records vector RMSE for the nine-input flood and
material models. The reported RMSE and reference noise floors divide those
values by $\sqrt d=3$ to obtain \eqref{eq:empirical-rmse}.

\subsection{Meeting and identity diagnostics}\label{app:experiment-diagnostics}
The four conditions are ordered as flood with $\eta=0.5,1.5$, followed by
material with $r_c=0.12,0.30$.

The saved production batches satisfy the samplewise telescoping identity
to floating-point precision. They also reproduce the decomposition of CU
into its matched finite contribution and correction. Meeting times pooled
over vectors and levels have means $2.006$, $2.019$, $2.012$, and $2.065$,
with observed maxima $6$, $15$, $6$, and $13$ in the same condition order.
These are per-level times, not maxima over all levels in a vector.
Nonzero vector corrections occur in $5.8\%$, $14.7\%$, $6.9\%$, and
$28.4\%$ of CU vectors. Short typical meeting times therefore do not make
the correction identically negligible.

As an additional check of the total-variance identity, we generate a separate joint-input draw $\bsY$, independently of the outer input and the MCMC simulation, and compare $A_d$ with $\{f(\bsX)-f(\bsY)\}^2/2$.
The paired differences between $A_d$ and
this quantity have standardized mean discrepancies $1.08$, $-2.02$,
$-1.74$, and $-0.40$. The second discrepancy is slightly beyond two estimated standard
errors. The largest absolute component discrepancy against the independent
reference is $2.28$ combined standard errors. These diagnostics check the
numerical implementation of the stated identities.

\section{Additional numerical results}\label{app:additional-results}

\subsection{Ising contributions and costs}
Full componentwise comparisons with enumerated Ising values are in
\cref{tab:ising-contributions}. Components and totals are rounded separately.

The response-count comparison at $N=8192$ is given in
\cref{tab:ising-results}. It includes the finite-time bias of matched
finite MCMC in its empirical RMSE, so the final column is not a
variance-only criterion for that biased method.

\begin{table}[htbp]
\caption{Dependent Ising inputs at $N=8192$, with $R=50$. The last column is the squared RMSE from \eqref{eq:empirical-rmse} multiplied by the mean number of response evaluations per experiment replication. It includes the finite-time bias of the finite-MCMC comparison.}\label{tab:ising-results}
\centering\small
\begin{tabular}{lrrr}
\toprule
Method & RMSE & Mean evaluations & RMSE$^2$ $\times$ evaluations\\
\midrule
CU & 0.01606 & 475,005 & 122.44 \\
IU & 0.02196 & 475,035 & 229.03 \\
Finite MCMC & 0.01965 & 466,944 & 180.23 \\
Conditional pick-freeze & 0.02229 & 73,728 & 36.64 \\
Nested variance & 0.02707 & 180,224 & 132.02 \\
\bottomrule
\end{tabular}
\end{table}

For $\beta=0.35$ and $N=8192$, the empirical pointwise interval
coverage averaged over the eight inputs is $94.75\%$ for CU and
$83.25\%$ for matched finite MCMC. Componentwise coverages have
substantial uncertainty at $R=50$; the eight inputs in a replication
are not eight independent experiment replications.
The fitted CU RMSE slope for $\beta=0$ is $-0.501$.

\subsection{Flood and material contributions}
The full flood and material CU estimates and standard errors are
shown in \cref{tab:flood-contributions} and \cref{tab:material-contributions}, respectively. For the flood model, the small negative estimate for
$L$ is consistent with a near-zero contribution and is not clipped.

\begin{table}[htbp]
\caption{CU estimates of the unnormalized flood contributions $\phi_j$,
with standard errors in parentheses below the estimates. All displayed
estimates, standard errors, and sums are multiplied by $10^3$.
Entries pool $R=30$ runs of $N=512$ completed vectors.
The two conditions correspond to different input laws.}
\label{tab:flood-contributions}\centering\small
\setlength{\tabcolsep}{3pt}
\begin{tabular}{@{}l*{10}{c}@{}}
\toprule
$\eta$ & $A_Q$ & $B_Q$ & $Q_1$ & $Q_2$ & $K_s$ & $Z_v$ & $Z_m$ & $L$ & $B_{\mathrm{width}}$ & Sum\\
\midrule
$0.5$ & $1.09$ & $1.15$ & $169.55$ & $172.34$ & $171.70$ & $226.52$ & $6.00$ & $-0.0093$ & $0.22$ & $748.56$\\
 & $(1.18)$ & $(1.26)$ & $(4.82)$ & $(4.90)$ & $(4.55)$ & $(4.74)$ & $(0.81)$ & $(0.0082)$ & $(0.13)$ & $(8.77)$\\
\midrule
$1.5$ & $6.55$ & $6.77$ & $157.25$ & $152.23$ & $165.38$ & $229.24$ & $4.05$ & $-0.0017$ & $0.13$ & $721.60$\\
 & $(2.32)$ & $(2.55)$ & $(5.39)$ & $(5.48)$ & $(4.55)$ & $(4.95)$ & $(0.86)$ & $(0.0081)$ & $(0.13)$ & $(8.50)$\\
\bottomrule
\end{tabular}
\end{table}

\begin{table}[htbp]
\caption{CU estimates of the unnormalized material contributions $\phi_a$
in squared response units, with standard errors in parentheses below the
estimates. Cells are indexed row by row on the $3\times3$ grid,
with cell 5 at the centre. Entries pool $R=30$ runs of $N=512$
completed vectors.}
\label{tab:material-contributions}\centering\small
\setlength{\tabcolsep}{2pt}
\begin{tabular}{@{}l*{10}{c}@{}}
\toprule
$r_c$ & 1 & 2 & 3 & 4 & 5 & 6 & 7 & 8 & 9 & Sum\\
\midrule
$0.12$ & $0.3329$ & $0.3731$ & $0.3633$ & $0.4200$ & $0.5054$ & $0.3895$ & $0.3236$ & $0.3778$ & $0.3134$ & $3.3990$\\
 & $(0.0154)$ & $(0.0164)$ & $(0.0155)$ & $(0.0167)$ & $(0.0170)$ & $(0.0153)$ & $(0.0153)$ & $(0.0163)$ & $(0.0145)$ & $(0.0371)$\\
\midrule
$0.30$ & $0.5355$ & $0.5785$ & $0.5654$ & $0.6198$ & $0.8325$ & $0.7058$ & $0.5390$ & $0.6270$ & $0.5398$ & $5.5432$\\
 & $(0.0285)$ & $(0.0287)$ & $(0.0279)$ & $(0.0296)$ & $(0.0316)$ & $(0.0300)$ & $(0.0260)$ & $(0.0276)$ & $(0.0261)$ & $(0.0588)$\\
\bottomrule
\end{tabular}
\end{table}

\subsection{Accuracy at fixed sample sizes}
\begin{figure}[!htbp]
\centering
\includegraphics[width=0.8\linewidth]{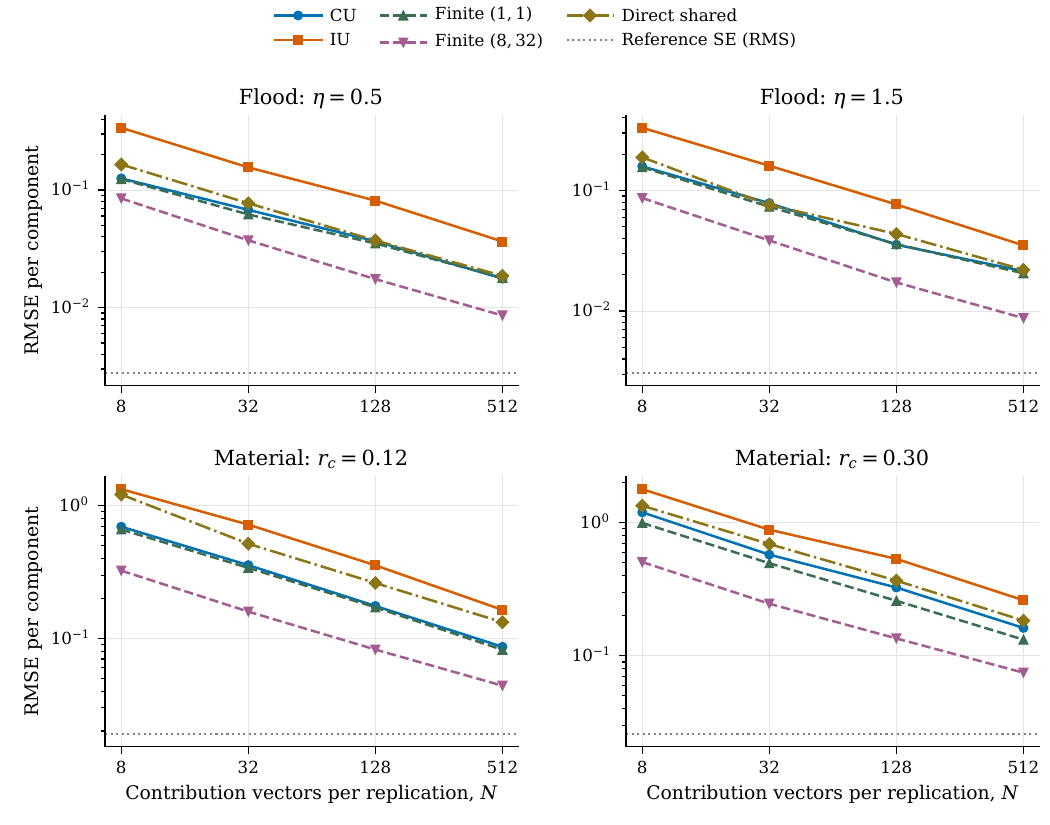}
\caption{Component-averaged RMSE against an independent finite Monte Carlo
reference, plotted against the number $N$ of completed contribution
vectors. Each curve uses prefixes $N=8,32,128,512$ of the same
$R=30$ replications. Horizontal dotted lines show the estimated
reference noise floor.}
\label{fig:new-accuracy}
\end{figure}

\Cref{fig:new-accuracy} compares the common-$N$ RMSE values for the
flood and material models. Each curve uses prefixes of the same independent experiment replications. The horizontal dotted line in each panel shows the estimated uncertainty of the Monte Carlo reference.

\FloatBarrier

\end{document}